\documentclass[10pt]{article}

\pdfoutput=1 %% recommended by arXiv (http://arxiv.org/help/submit_tex#pdflatex)

\usepackage{lipsum} % Package to generate dummy text throughout this template

\usepackage[sc]{mathpazo} % Use the Palatino font
\usepackage[T1]{fontenc} % Use 8-bit encoding that has 256 glyphs
\usepackage{microtype} % Slightly tweak font spacing for aesthetics

\usepackage[hmarginratio=1:1,top=32mm,columnsep=20pt]{geometry} % Document margins
\usepackage{multicol} % Used for the two-column layout of the document
\usepackage[hang, small,labelfont=bf,up,textfont=it,up]{caption} % Custom captions under/above floats in tables or figures
\usepackage{booktabs} % Horizontal rules in tables
\usepackage{float} % Required for tables and figures in the multi-column environment - they need to be placed in specific locations with the [H] (e.g. \begin{table}[H])
\usepackage{hyperref} % For hyperlinks in the PDF

\usepackage{lettrine} % The lettrine is the first enlarged letter at the beginning of the text
\usepackage{paralist} % Used for the compactitem environment which makes bullet points with less space between them

\usepackage{abstract} % Allows abstract customization
\usepackage{titlesec} % Allows customization of titles
\renewcommand\thesection{\Roman{section}}
\titleformat{\section}[block]{\large\scshape\centering}{\thesection.}{1em}{} % Change the look of the section titles

\usepackage{fancyhdr} % Headers and footers
\usepackage[pdftex]{graphicx}
\usepackage{amsmath}
\usepackage{algpseudocode}

\newcommand{\OMIT}[1]{}%!TEX TS-program =  

\newcommand{\coordUpdate}{\texttt{coordUpdate}}

\OMIT{
\title{\vspace{-5mm}\fontsize{16pt}{10pt}\selectfont\textbf{
On Combinatorial Optimization and Self-Avoiding Walks \\[0.4ex]
in Hyperhedra: from Minimum Vertex Cover \\[0.4ex]
and Maximum Clique to Protein Folding
}} % Article title
}
\title{\vspace{-5mm}\fontsize{16pt}{10pt}\selectfont\textbf{
On Self-Avoiding Walks across n-Dimensional Dice  \\[0.5ex] 
and Combinatorial Optimization: An Introduction
}} % Article title

\author{
\large
\textsc{Franc Brglez}%\thanks{A thank you or further information}
\\[2mm] % Your name
\normalsize Computer Science \\
\normalsize NC State University, Raleigh, NC 27695, USA \\ % Your institution
\normalsize \href{mailto:brglez@ncsu.edu}{brglez@ncsu.edu} % Your email address
\vspace{-5mm}
}
\date{}

\begin{document}

\maketitle % Insert title

\thispagestyle{fancy} % All pages have headers and footers
%\thispagestyle{empty} % this removes page number on title page!!!!
%----------------------------------------------------------------------------------------
%	ABSTRACT
%----------------------------------------------------------------------------------------

%\begin{abstract}
\vspace*{-0.25ex}
\noindent %\lipsum[1] % Dummy abstract text
\begin{center}
\begin{minipage}[]{0.912\textwidth}
{\large\bf Abstract~--~}
{\bf 
Self-avoiding walks (SAWs) were introduced in chemistry
to model the real-life behavior of chain-like entities such as solvents and polymers, 
whose physical volume prohibits multiple occupation of the same spatial point.
In mathematics, a SAW lives in the n-dimensional lattice $\mathbb{Z}^n$  which consists
of the points in $\mathbb{R}^n$  whose components are integers.

~~~~In this paper, SAWs are a metaphor for walks across {\em faces} of $n$-dimensional dice,
or more formally, a {\em hyperhedron} family ${\cal H}(\Theta, b, n)$.
Each face is assigned a label 
\{${\underline \varsigma}:\Theta({\underline \varsigma})$\};
${\underline \varsigma}$ represents a unique $n$-dimensional coordinate string ${\underline \varsigma}$, 
~$\Theta({\underline \varsigma})$ is the value of the function $\Theta$ for ${\underline \varsigma}$.
The walk searches $\Theta({\underline \varsigma})$ for optima
by following five  simple rules:
(1) select a random coordinate and mark it as the `initial pivot';
(2) probe all unmarked adjacent coordinates, then select and mark the coordinate with the 'best value' as the new pivot; 
(3) continue the walk until either the 'best value' <= `target value' or the walk is being blocked by adjacent coordinates that are already pivots;
(4) if the walk is blocked, restart the walk from a randomly selected `new initial pivot';
(5) if needed, manage the memory overflow with  a streaming-like buffer of appropriate size. 
Hard instances from a number of problem domains, including the 2D protein folding problem, with up to $(2^{25})*(3^{24})$ coordinates, have been solved with SAWs in less than 1,000,000 steps -- while also exceeding the quality of best known solutions to date.
}

\end{minipage}
\vspace*{4ex}
\end{center}
%\end{abstract}

%----------------------------------------------------------------------------------------
%	ARTICLE CONTENTS
%----------------------------------------------------------------------------------------

\begin{multicols}{2} % Two-column layout throughout the main article text

\section{Introduction}
\label{sec_introduction}
\noindent
Instances of combinatorial problems arise in many contexts such as operations research, %automated reasoning, 
computer-aided design,
% and manufacturing,  
machine learning, robotics, data mining, 
%computer networking, 
bioinformatics, etc. 
An exhaustive search for an optimum solution is not possible for
most instances of practical size due to the huge number of feasible solutions.
The question arises about the choice of heuristic algorithms 
to be deployed by the solver. To date, 
stochastic search methods offer the best compromise,
including Metropolis-Hastings algorithm~\cite{1953-JChPh-Metropolis,1970-MCMC-Hastings}, simulated annealing~\cite{1983-Science-Kirkpatrick,2011-wikipedia-Simulated_annealing}, Gibbs sampling~\cite{1984-Gibbs_sampling-Geman}, 
tabu search~\cite{1989-TabuSearch-Glover,1990-TabuSearch-Glover}, and 
many others. New heuristics are emerging on Wikipedia and in journals under metaphors
such as
ant colonies, %~\cite{2011-Wikipedia-Ant_colony},
bird flocks,
%particle swarms, %~\cite{2011-wikipedia-Particle_swarm_optimization},
natural disasters, %~\cite{2011-wikipedia-Great_Deluge_algorithm},
biological processes, etc. %~\cite{2011-wikipedia-Evolutionary_computation}, etc.

Our approach is simple; we only take a few liberties with rigorous mathematical notation. 
When we refer to a {\em function} $f(x_1^i, x_2^i, \ldots x_n^i)$, we imply 
an {\em objective function}, which in general is a {\em multivalued function},
returning a {\em value} for a specific {\em coordinate} $(x_1^i, x_2^i, \ldots x_n^i)$.
The {\em support set} of the function is defined in terms of such coordinates.
A combinatorial problem is defined by its function {\em and} its {\em coordinate type}.
Coordinates are represented as a set of strings, such as $01011...$ for a binary coordinate,
$210210...$ for a ternary coordinate, $4,2,5,3, ...$ for a permutation coordinate, etc.
A combinatorial problem can also be stated in terms of 
{\em concatenated} coordinates of different types. For example, we define the 2D protein folding problem {\em on a square lattice} by computing its function values with coordinates represented as {\em a concatenation of binary and ternary coordinates}. 

We define
a {\em walk} as a sequence of steps that chain a set of 
{\em pivot coordinates}, {\em adjacent coordinates} as the
{\em local neighborhood} of the pivot coordinate, and 
{\em probing} as evaluating the function for values in this neighborhood.
A {\em feasible solution} of the combinatorial problem is a pair 
{\em (coordinatePivot}{\bf :}{\em valuePivot)}. Once we capture the description of the combinatorial problem in these terms, 
the search procedure is as simple and as generic as the five rules outlined in the abstract -- for {\em any} combinatorial problem. For more about this notation and
examples of various problem instances, 
see~\cite{Lib-OPUS2-walk-2011-EV-Brglez}.

We have a number of on-going projects with the goal to prototype
SAWs as a powerful general-purpose search procedure that,
unlike alternatives, does not require knobs and dials to set-up.
These projects include instances of
problems defined for Golomb rulers, integer partitioning, maximum independent set, minimum vertex cover and maximum clique, graph linear arrangement, job scheduling, etc. A nearly completed project
demonstrates significant improvements in 
solutions of the notoriously hard 
{\em labs problem}~\cite{Lib-OPUS2-labs-2013-arxiv-Boskovic}: here we compare, side-by-side, the 
performance of state-of-the-art memetic/tabu and SAW solvers.
In the present paper we apply SAW to solve the 2D protein folding problem 
{\em on a square lattice}~\cite{Lib-OPUS-bioinfo.fold-2009-InfSyst-Istrail-survey}.
Since this implementation is based on a scripting language, it is
not suitable for solving very large problems. However, the solver does achieve 
a number of state-of-the art solutions on a significant subset of problem instances from the literature and an
asymptotic performance that may well dominate alternative solvers when fully implemented.

The paper is organized as follows. 
To motivate the approach taken in this paper, Section \ref{sec_motivation} starts with a fable about
Gretel and Hansel who devise distinctive methods to search for a {\em pass-key}.
% fg_BT_dice ; fg_BT_dice_walks
% 
Section III %\ref{sec_notation} 
% fg_lattice_saw
% fg_BT_conformations_intro 
% fg_BT_neighborhood_calc
% fg_global_search
formulates the problem and concludes
with pseudo code, describing global search with self-avoiding walk.
Section IV %\ref{sec_experiments} 
% fg_snake_spiral
% tb_BT_conformations
% fg_BT_conformations
summarizes a number of performance experiments in
2D protein folding problem defined {\em on a square lattice}.
With some 1000 independent solutions for each member of 10 
instance classes of increasing size (with at least 3 instances in each class), 
these  experiments  not only replicate
the earlier work of others, but also reveal
new and improved folding conformations.
The paper concludes with directions for future work.

\section{Motivation}
\label{sec_motivation}
% fg_BT_dice ; fg_BT_dice_walks
We introduce a fable to motivate our approach.
It involves Gretel and Hansel, living in two adjacent apartments, and a Joker
whose omnipresence is never revealed directly.
Gretel and Hansel are returning from a party. They discover not only that locks have been changed on both apartment doors with punch-key locks but also that mats that hid the keys were replaced with two urns, each containing a set 36 tickets. Each ticket has a printed label with five digits in the format xx.yy:z; only one label opens Gretel's door, and only one opens Hansel's door. The two sets are identical.

Who gets in first? Watching Gretel, she takes
the ticket from the urn and if she does not succeed in opening the door, she puts the ticket into her handbag and retrieves another ticket. Hansel, who had a few drinks at the party, takes the ticket and if he does not succeed in opening the door, returns the
ticket to the urn. We  say that Gretel is sampling contents of the urn without replacement, while Hansel is sampling with replacement. 
The probability of Gretel finding the correct ticket on trial $k$ follows uniform distribution, given $n$ tickets:
the probability is $1/n$, the mean value is $(n+1)/2$, and the variance is $(n^2 -1)/12$.
However, the probability of Hansel finding the correct ticket on trial $k$ follows geometric distribution:
the probability is  $(1/n)(1-(1/n))^{k-1}$, the mean value is $n$,  and the variance is $n^2(1-(1/n))$. The point of the fable so far: we learn that in a search scenarios such as described here,
one can improve the chance of {\em first success} by dynamically reducing the search space
after each trial. In the best case for Gretel, the capacity of her handbag must match the capacity of the urn. If the capacity of the handbag is
less than the capacity of the urn, and the handbag gets full before finding the key,
she needs to return the unsuccessful ticket to the urn before proceeding with the next trial. In the average case, Gretel's search, even with handbag of limited capacity,  always requires fewer trials than Hansel's. 

\begin{figure*}[]
%\small
\begin{center}
\vspace*{-1ex}
\hspace*{-3.1ex}
\begin{tabular}{p{0.45\textwidth}p{0.07\textwidth}p{0.53\textwidth}}
 
   \begin{tabular}{p{0.40\textwidth}}
   %\hspace*{-2.0ex} \vspace*{-2.0ex}
   \includegraphics[width=0.45\textwidth]{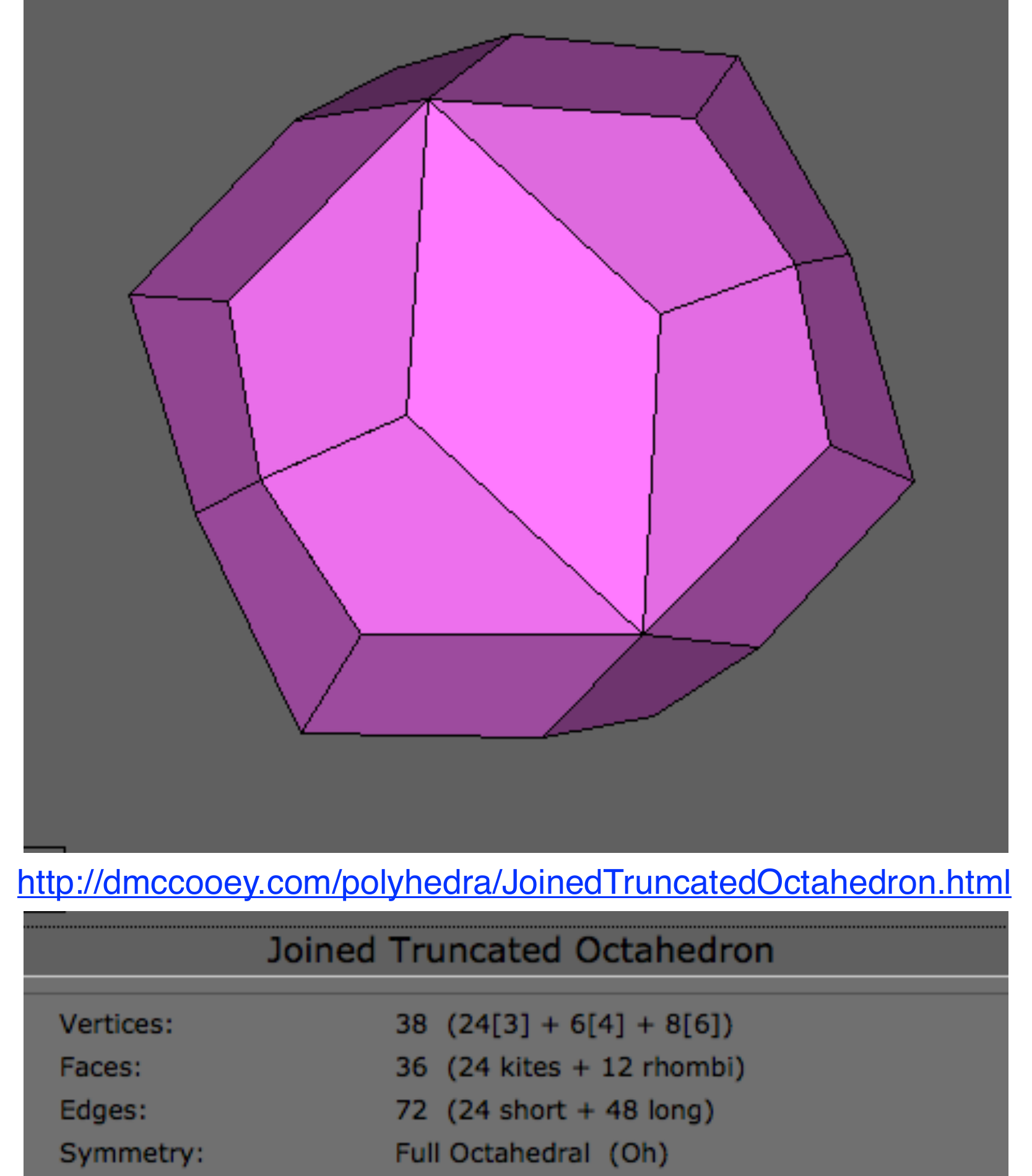}
   \end{tabular}
&
\begin{tabular}{p{0.07\textwidth}}
\par
\end{tabular}
&
\hspace*{-10.5ex}
%\fbox{ 
\begin{tabular}{p{0.53\textwidth}}
   %\par\vspace*{1ex}
   %\ragged-right
%\lipsum[1]
\small
The item on the left is a polyhedron with 36 faces and 72 
edges~\cite{2013-Web-McCooey-polyhedra}. 
Each face has 4 adjoining faces. This polyhedron is an approximation of the a virtual combinatorial object, a hyperhedron introduced next. By assigning to each face a
unique coordinate as a concatenation of a binary strings of length 2 and a ternary string of length 2, we create a hyperhedron with $2^2\times3^2 = 36$ faces, the same as polyhedron. However, this hyperhedron has 84 edges compared to 72 in the polyhedron. We count the edges by creating a Hasse 
graph~\cite{Lib-OPUS2-walk-2011-EV-Brglez}: each face is assigned a vertex with a unique label and the edges between vertices represent adjacencies between  
faces. We find that the number of edges between vertices varies from 4 to 6.

~~~The label always contains a unique coordinate string,
and in most cases, the label is extended with {\em a value} returned 
by the function evaluated with the coordinate. The Hasse graph is drawn as an {\em undirected} layered graph on a grid such as the one below: it has 36 vertices and 84 edges with labels such as 00:00:2, 00.10:9, 
and 01:21:9; the string following ':' represents the value. 
We say that vertices 00.00:2 and 00.10:9 are adjacent
since the distance between coordinates is 1, while
coordinates 00.10:9 and 01:21:9 are not adjacent since the distance
is 3.
\end{tabular}
%}
\end{tabular}

\par\vspace*{-8ex} %\\[-1.75ex]
\hspace*{-0.0ex}
\includegraphics[width=0.95\textwidth]{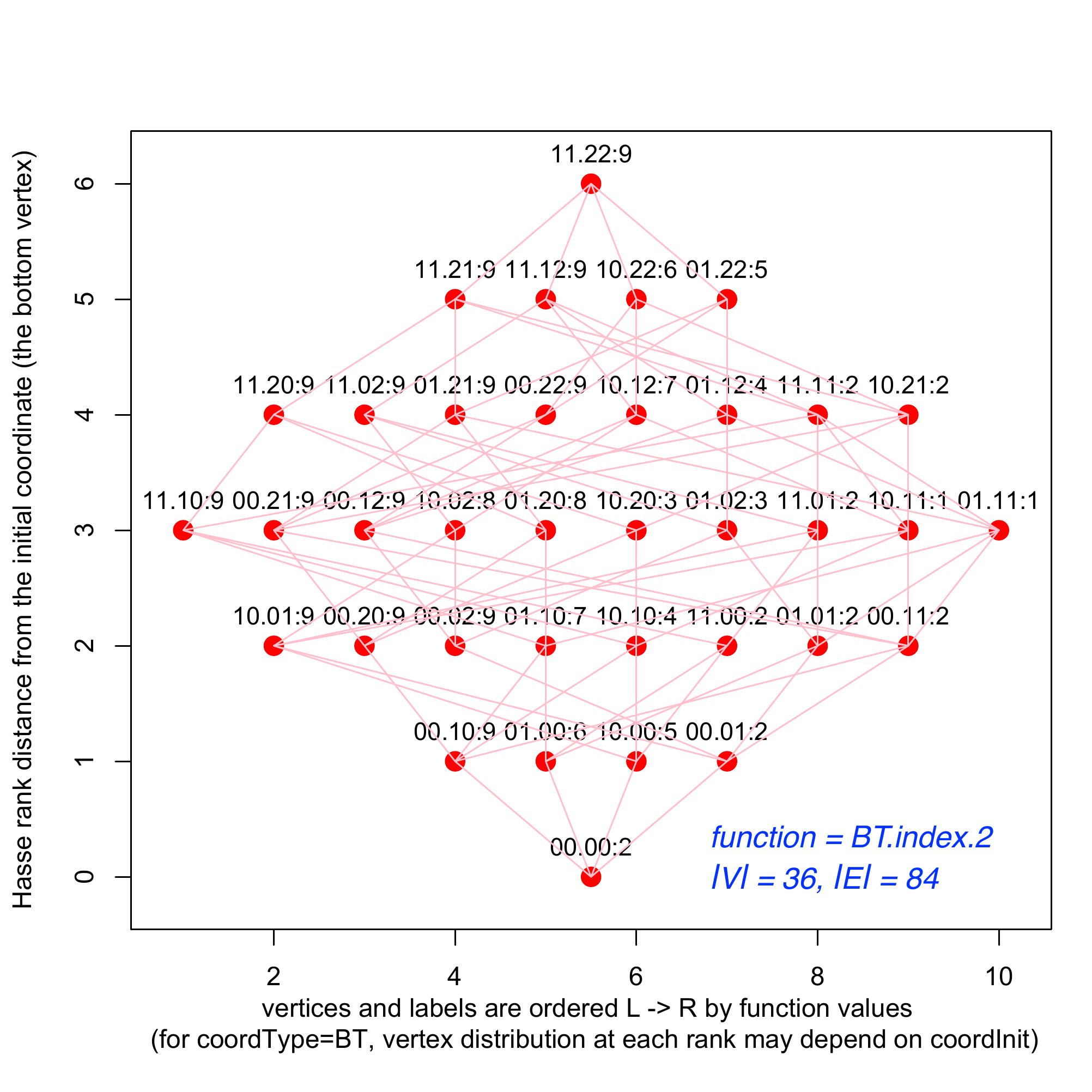}
\vspace*{-2ex}
\caption{
A polyhedron solid and a hyperhedron projection: each has 36 faces, but face-to-face  adjacencies are different.
}
\label{fg_BT_dice}
%\vspace*{-6ex}
\end{center}
\end{figure*}

Another surprise awaits after Gretel and Hansel manage to make an entry. 
Neither has stepped into their apartment's vestibule; instead, each is now standing on  
a four-sided platform (in their own apartment) and 
can see, besides the platform on which they are standing, only
the surfaces of the four adjacent platforms sloping downwards. 
Each of them  believes that she/he is standing on a 
face of a huge platonic solid, such as the polyhedron with 36 faces 
and 72 edges between the faces shown in
in Figure~\ref{fg_BT_dice}.
%just as they remember from
%their class in crystallography. 
Neither realizes
that they stepped into a virtual world where not everything is as it seems.
The face on which they are standing belongs to a
virtual combinatorial object {\em hyperhedron}, also with 36 faces,
but as they will walk from face to face, they will discover that some faces have five adjacent faces, some have even six.

%\begin{figure*}[ht!!] % this does work, i.e. [ht!!] (need to leave at least space of 3 lines at the bottom)
\begin{figure*}[] % [] seems to work best ; this does NOT work [H]
\begin{center}
\vspace*{-6.5ex}
\hspace*{-2.5em}
\includegraphics[width=1.05\textwidth]{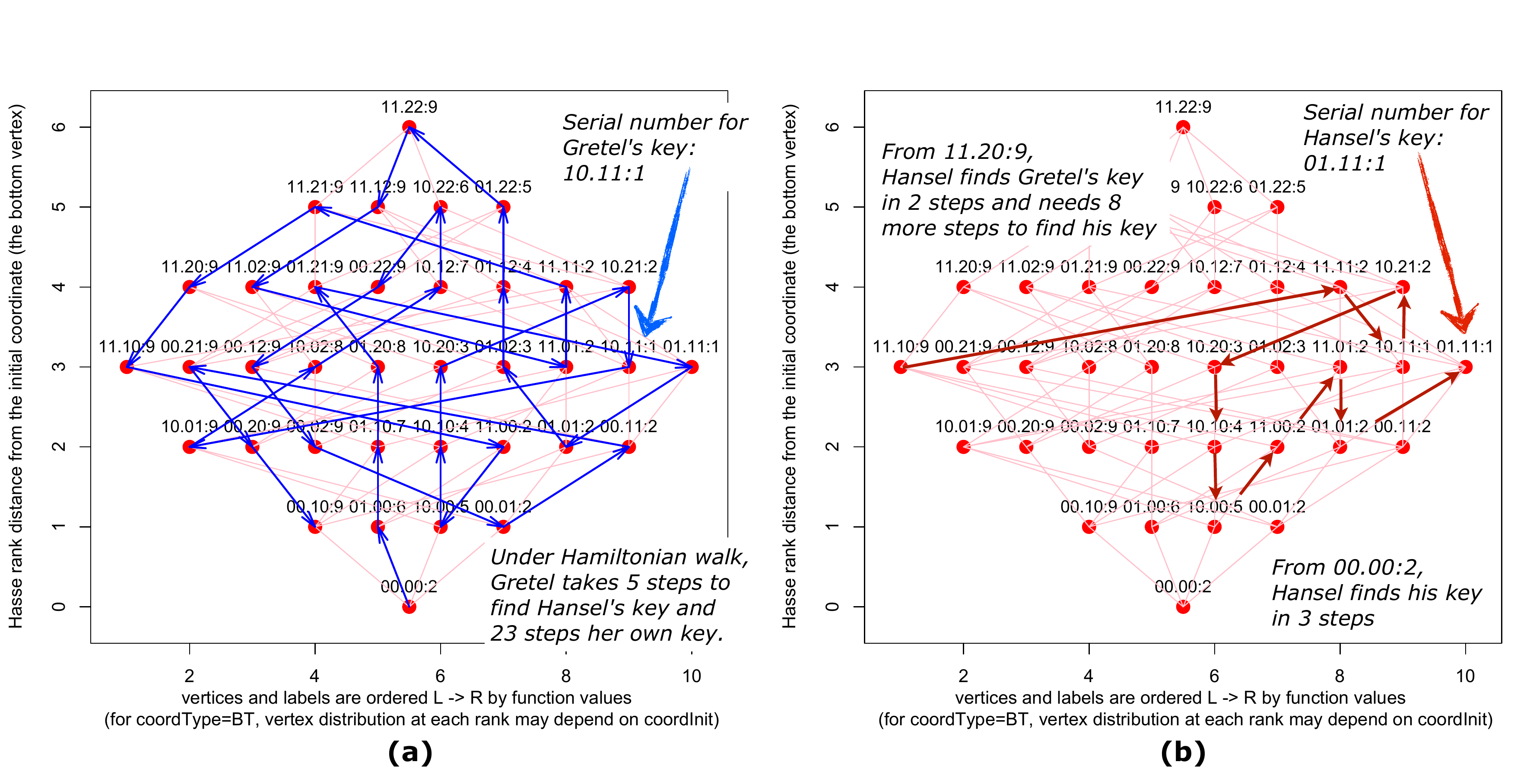}
\vspace*{-3.5ex}
\caption{
A Hamiltonian walk in the Hasse graph taken by Gretel and a 
self-avoiding walk taken Hansel. 
Vertices traversed during the walk are in two categories:
(1) only binary coordinate is changing, ternary coordinate is fixed;
(2) binary coordinate is fixed, only ternary coordinate is changing.
At each pivot, before Hansel decides on the next step, he probes the 
function values at all adjacent coordinates (that are not yet pivots in the walk) and chooses the coordinate with
'valueBest'. If there are multiple {\em adjacent coordinates} with the same 'valueBest',
the choice is random.
Gretel's walk, self-avoiding by definition, continues until 'valueBest = valueTarget = 1' and the key found fits her lock,
Hansel's walk terminates when the key found fits his lock.
}
\label{fg_BT_dice_walks}
\end{center}
\vspace*{-4.0ex}
\end{figure*}

Joker has replaced the two urns with two hyperhedrons and  pasted the tickets from each urn into the center of the face in each the hyperhedron, with labels showing. He also hid Gretel's pass-key under one ticket, and Hansel's key under another ticket. 

Only Joker has the global view of the hyperhedron. He  
interprets it as follows. He moves {\em inside}
the hyperhedron, finds the center of the face, and attaches one end of a string to the center and attaches
the other string to the center of the adjacent face. He repeats the process for all faces and thus creates a graph; a graph with 36 face-centered vertices and 84 edges. To represent this graph in the plane,
he defines {\em a distance  between the coordinates} assigned to each label and makes a projection of the graph
as a layered graph shown in the bottom of Figure~\ref{fg_BT_dice}. This graph is not visible to Gretel and Hansel, however, the graph enables Joker to trace each step they would make during their search.
Joker also assigned function values to each coordinate: his choice of values
is expected to confound Gretel and Hansel in their search.
He gives both one, {\em and only one}, hint about the pass-key: the ticket most likely hiding the pass-key is the one where {\em value} on the label is 1 or less than 1. If Gretel find Hansel's key first, the key would not fit and she needs to continue the search -- and vice versa for Hansel.

Who will find the pass-key to the apartment first?
Each is standing on the face with the label 00.00:2
(at the bottom of Figure \ref{fg_BT_dice}). From this face,
Gretel and Hansel can see only the four
adjacent faces: 00.10:9, 01.00:6, 10.00:5, 00.01:2. Their task is to {\em walk} from face to face
until they find the {\em pass-key} to their old apartment. 
%The presence of the pass-key is
%revealed only after probing the pasted ticket with a single knock.

Gretel is a computer science major and remembers a lecture about 
Hamiltonian walks in graphs. She knows that she is standing on one of 36 faces and that
if she associates each face with a vertex in a graph and 
the edges between adjacent faces with edges in this graph,
she can {\em compute and remember} the path that  visits each face only once.
In the worst case, she will take 35 steps to find the key.
The procedure she uses to compute the coordinates for each step in the Hamiltonian walk
is not as simple to explain as the procedure used by Hansel and explained next.
Suffice it to say that function values associated with each coordinate 
have no role in determining the Hamiltonian path in the graph.
An example of Gretel's walk,
as traced by Joker, is shown in \ref{fg_BT_dice_walks}-a.
She takes 5 steps to find Hansel's key and needs to continue
for a total of 23 steps before finding her key.
The first step, from 00.00:2 to 00.00:6, is a deliberate step 
in this Hamiltonian walk -- a step that Hansel would never have taken from
this starting position, for reasons we explain next. 

Hansel's major is land surveying and he takes the hint about function values associated with each coordinate very seriously. He devises a few rules before starting the walk:
(1) mark the face from where the walk starts with an easy-to-spot token; later on in the paper, we call this
face the initial pivot.
(2) probe each adjacent face that has not yet been marked and write its value on a list.
(3) select the adjacent face with the smallest value, step on this face, 
call it a current pivot, and mark it with a new token.
If there are several faces with the same value, make a random selection.
(4) repeat step (2) from the current pivot until reaching the target value.
The process of marking the pivots during the walk with tokens makes this walk self-avoiding.
Hansel can run into a problem with these rules in two cases:
(1) he runs out of tokens and can no longer mark the walk;
(2) he steps onto a face where all adjacent faces have been marked already, i.e. the walk is trapped.
In either of these cases, Hansel has to collect all tokens and restart the walk from a new
face, now selected by a random jump. 
An example of Hansel's walk, as traced by Joker,
is shown in Figure \ref{fg_BT_dice_walks}-b. 
%The caption in this figure uses some of the terms not yet formally defined by Hansel, 
%they are defined in the section that follows. 
While Hansel can find the ticket that hides his pass-key in 3 steps or less
from many initial positions, it takes 10 steps to find his key
if he starts from 10.10:9 and takes the third step to 10.21:2 instead of 00.11:2
(both of these choice are equally likely).

What have we learned from the second part of the fable is this:
(1) A Hamiltonian walk, while self-avoiding by definition, should not be the first choice under the search scenarios described in this paper. Moreover, the approach would not scale to large problem instances. 
(2) On the other hand, rules devised by Hansel seem to be highly effective. 
The good news is that these rules are now enabling 
effective combinatorial searches not only in the cases of protein folding investigated in this paper but also on hard combinatorial problems in other domains, notably the low autocorrelation binary sequence problem, 
where the self-avoiding walks solve large problems that could not be solved with
state-of-the art memetic/tabu search 
methods~\cite{Lib-OPUS2-labs-2013-arxiv-Boskovic}.
Moreover, problem of self-avoiding walks getting trapped has not presented itself neither in the case of protein folding nor the case of the labs 
problem~\cite{Lib-OPUS2-labs-2013-arxiv-Boskovic}. 

We used to call these walks {\em Hansel's walks}
until we learned about  polymer and protein chain folding and 
{\em self-avoiding walks}~\cite{Lib-OPUS-walk-self-avoiding-2013-Wikipedia}. 
In our context, the self-avoiding walks
are walks in hyperhedra, a virtual world, not in a space of real-world constraints imposed by various lattice formulations in two or three 
dimensions~\cite{Lib-OPUS-walk-self-avoiding-1984-JChemPhysics-Hemmer}. 
In our formulation we deal with real-world folding constraints by way of computing the 
function values in terms of our coordinate system which foremost defines positions
and distances between face-centered vertices in hyperhedra.
For problems such as protein folding, some  coordinates may induce a penalty value when a conflict
is detected during folding;
the penalty value assigned may depend on the perceived level of conflict. Hansel's strategy, of {\em always} selecting the best {\em available} value in the local neighborhood for the next step, keeps the walk going, {\em across faces of a specific hyperhedron}, for as long as required.

% What does appear magical at this point is the fact that our SAWs consistently find the target value {\em 
% before} the walk gets trapped due to congestion of adjacent pivots
% in the hyperhedron. If the function always returned a constant value, the SAW would select
% each step randomly -- which would most likely entrap the walk before 
% it could ever complete the Hamiltonian walk.  

\section{Notation and Definitions}
\label{sec_notation}
%
% fg_lattice_saw
% fg_BT_conformations_intro 
% tb_BT_conformations_plans
% fg_BT_neighborhood_calc
% fg_global_search
Self-avoiding walks (SAWs) were first introduced by the chemist Paul Flory
in order to model the real-life behavior of chain-like entities such as solvents and polymers, 
whose physical volume prohibits multiple occupation of the same spatial point
\cite{Lib-OPUS-walk-self-avoiding-2013-Wikipedia}.
In mathematics, a SAW lives in the n-dimensional lattice $\mathbb{Z}^n$ which consists
of the points in $\mathbb{R}^n$ whose components are all integers 
\cite{Lib-OPUS-walk-self-avoiding-1994-Springer-Slade}. 

In Section \ref{sec_motivation}, we illustrated a grid of a {\em finite dimension} that was created by projecting 
face-centered vertices in  a hyperhedron, onto a plane as a Hasse graph. This section illustrates: 
(1) projections of vertices in Hasse graphs that have 1-to-1 relationship to lattices defined by {\em unit cells};
(2) example of a SAW-in-a-hyperhedron search for best folding of a protein chain of size $n$ on a specific 2D lattice ;
(3) formalized definitions of walks and a generic SAW pseudo-code as a principal component of a global stochastic search solver.

\par
\begin{figure*}[] % [] seems to work best ; this does NOT work [H]
\begin{center}
\vspace*{-4.5ex}
%\hspace*{-3.0ex}
\includegraphics[width=0.795\textwidth]{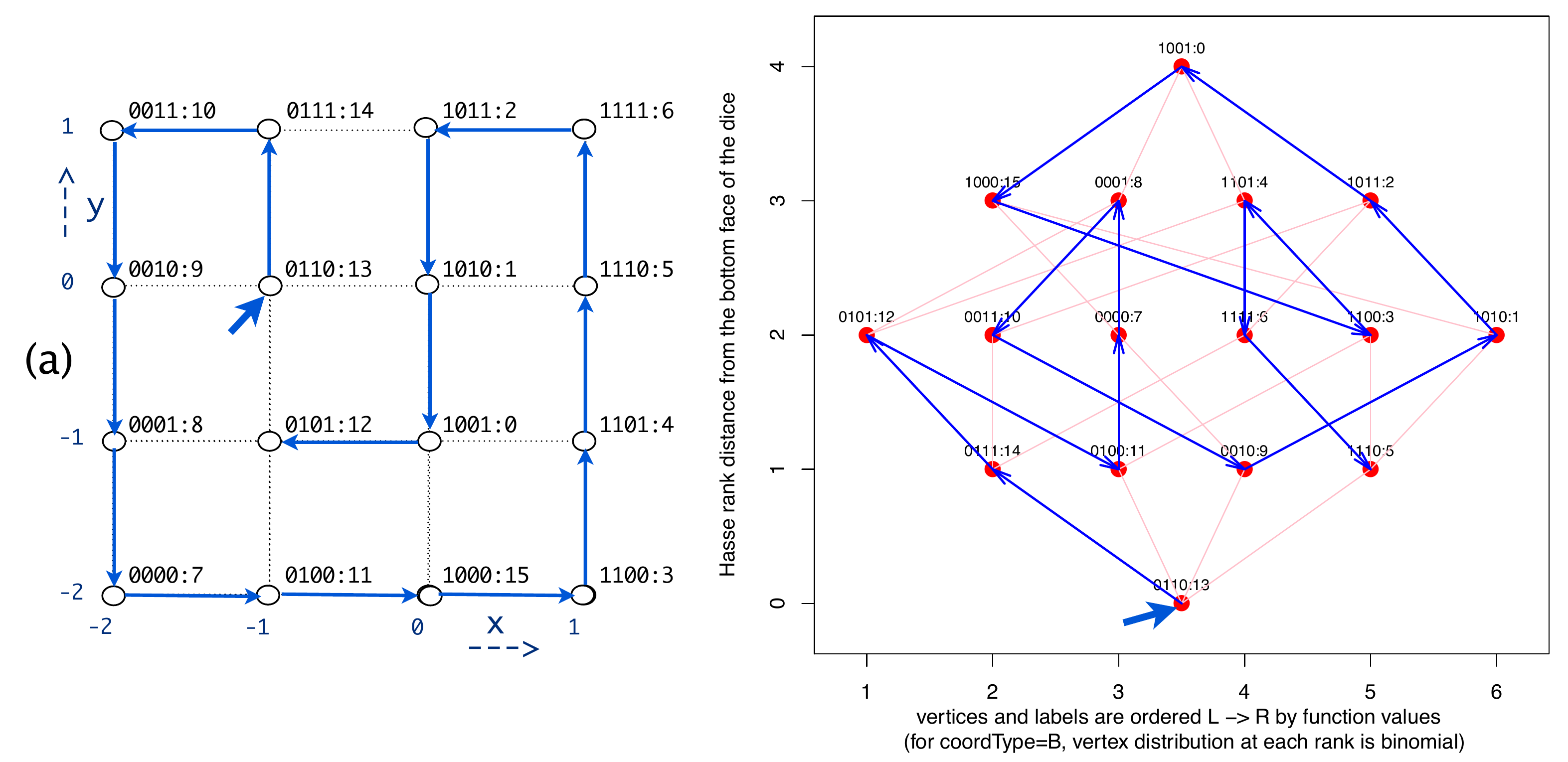}
\\[0.5ex]
\includegraphics[width=0.795\textwidth]{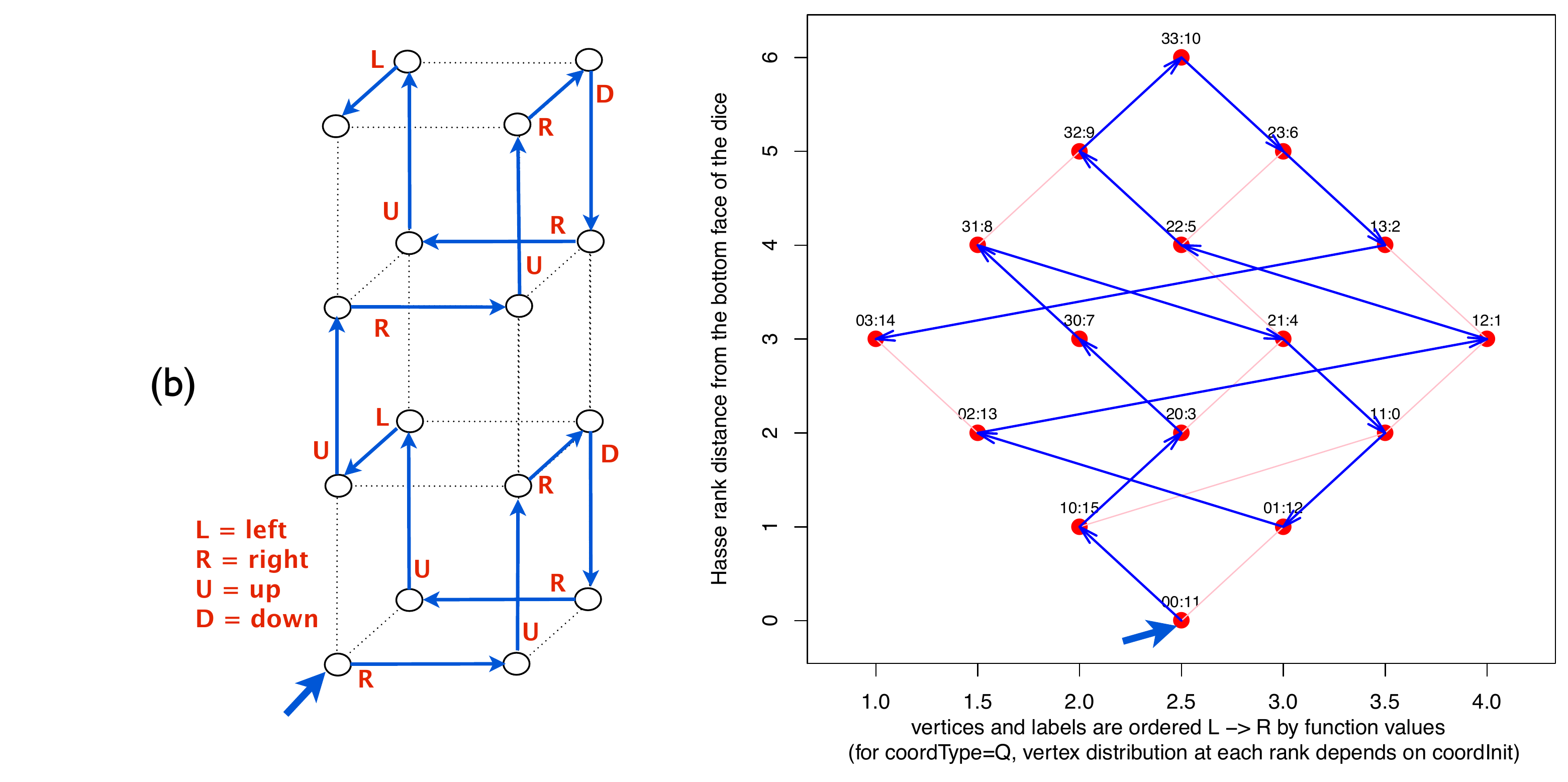}
\\[0.5ex]
\includegraphics[width=0.795\textwidth]{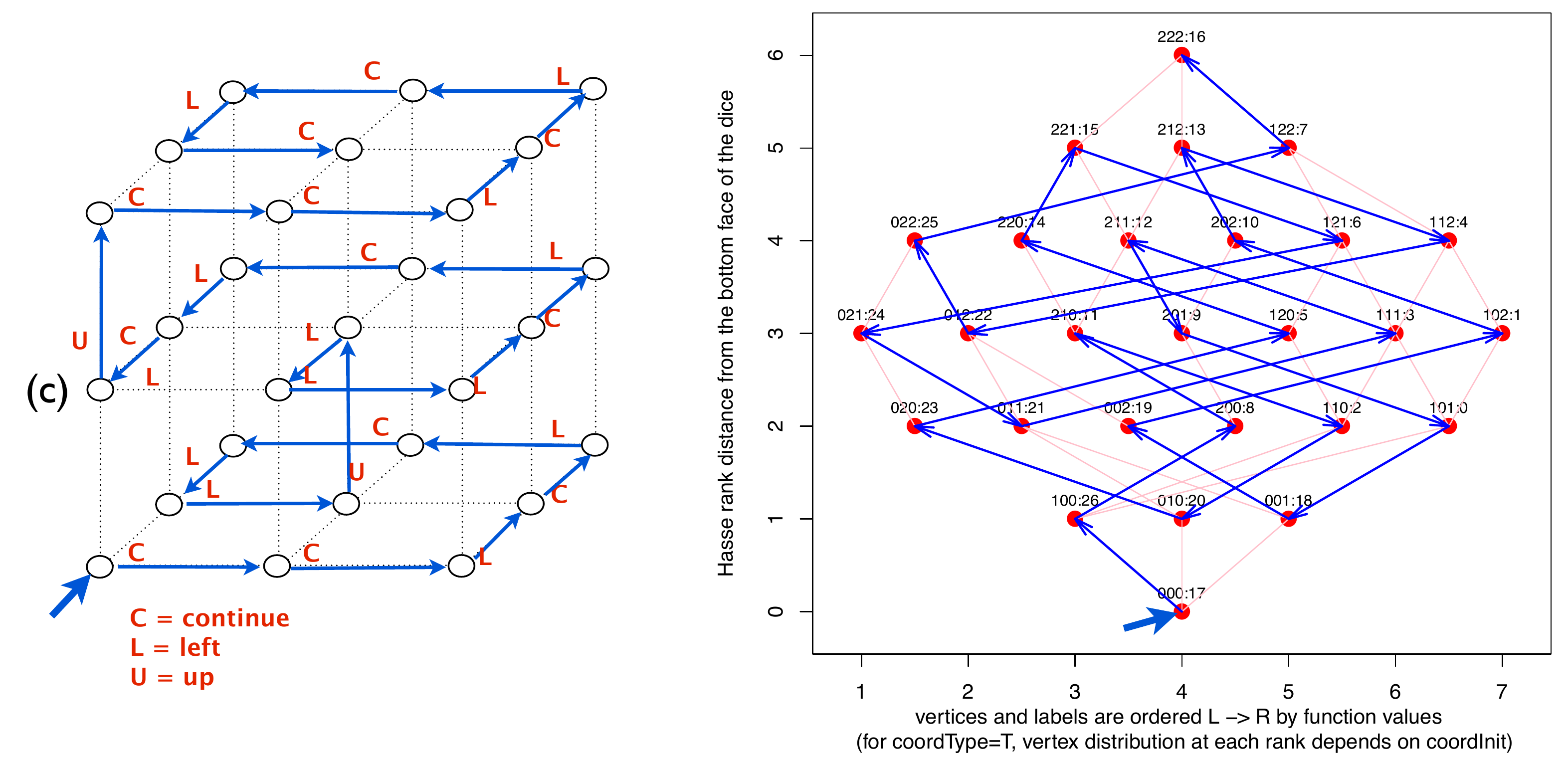}
\vspace*{0ex}
\caption{
Two sides of the same coin:
instances of three self-avoiding walks of lengths
$2^4 - 1$, $4^2 - 1$, and $3^3 -1$ in 2-D and 3-D in
{\bf unit cells}, subsets of points on a grid in 
a lattice~\cite{Lib-OPUS-chemistry-1970-Dover-Pauling-book},
versus instances of
three Hamiltonian walks of the same length in 
{\bf Hasse graphs}~\cite{Lib-OPUS2-walk-2011-EV-Brglez}
defined by dimensions 
4 (wrt to base 2), 2 (wrt to base 4), and 3 (wrt to base 3). The 
walks in unit cells are contiguous only with respect to coordinates defined in lattices.
Similarly, the walks in Hasse graphs are contiguous only with respect to coordinates defined  in Hasse graphs.
}
\label{fg_lattice_saw}
\end{center}
\vspace*{-4.0ex}
\end{figure*}

\vspace*{1ex}\noindent
{\bf Lattices, unit cells, and graphs.}
A lattice is a periodic array of points on a grid in space.
A {\em unit cell} is
a subset of $|V|$ points on a grid in 
a lattice~\cite{Lib-OPUS-chemistry-1970-Dover-Pauling-book}.
A self-avoiding walk in a unit cell and a
Hamiltonian walk in a {\em Hasse graph} with $|V|$ vertices
\cite{Lib-OPUS2-walk-2011-EV-Brglez}
can be considered as two faces of the same coin\footnote{
We are making this point metaphorically: 
a unit cell is a specific subset of grid points in a lattice 
\cite{Lib-OPUS-chemistry-1970-Dover-Pauling-book} 
and details about crystal structure arrangements and unit cells are
a far beyond the scope of this paper.}.
We illustrate this premise with the three examples in Figure \ref{fg_lattice_saw}.
In Figure \ref{fg_lattice_saw}-a on the left, the primitive cell is a square, forming a unit cell of 9 squares
with 16 grid points. On the right, we have a Hasse graph with 16 vertices with binary coordinate labels; this graph is regular since
the degree of each vertex is 4 -- i.e. each vertex has 4 immediate neighbors. 
% http://en.wikipedia.org/wiki/Regular_graph
% In graph theory, a regular graph is a graph where each vertex has the same number of neighbors; i.e. every vertex has the same degree or 
% valency. A regular directed graph must also satisfy the stronger condition that the indegree and outdegree of each vertex are equal to each 
% other.[1] A regular graph with vertices of degree k is called a k?regular graph or regular graph of degree k.
Given the starting point in the unit cell,
we can express the walk in terms of directional encoding (North, South, East, West): for the first six steps we would write NWSSSE.
Given the starting point in the graph, we express the walk as a sequence of its pivot coordinates (\ref{eq_coord_pivots}): {\sf 0110, 0111, 0101, 0100, ...} etc.
However, there is a significant difference in the two data structures: in the unit cell, 
neighborhood size, depending on the location in the grid,  varies from  2, 3, to 4,
whereas in the graph, each vertex has 4 neighbors.
%\begin{figure*}[ht!!] % this does work, i.e. [ht!!] (need to leave at least space of 3 lines at the bottom)

The crux in drawing the Hasse graph into its distinct layers is the notion of {\em Hasse rank distance} between the
vertices with respect to the reference vertex (or the origin vertex) placed at the very bottom of the 
graph~\cite{Lib-OPUS2-walk-2011-EV-Brglez}. When coordinates are binary strings, the rank distance is the familiar {\em Hamming distance}, for coordinate that represent permutations, the rank distance is the {\em permutation inversion distance}, the rank distance between the ternary and quaternary coordinates in Figure~\ref{fg_lattice_saw}
is defined as an arithmetic addition of modulo-3 or modulo-4 distances between coordinate components.
For example, the distance between 2101 and 0201 is 2+1+0+0=3, the distance between  3210 and 0123 is 3+1+1+3=8, etc. The distance between two coordinates that have been concatenated, shown in the Hasse graph in
Figure~\ref{fg_BT_dice}, is an arithmetic addition of distances between the corresponding concatenated segments, 
for example the distance between 00.10 and 01.21 is (0+1)+(1+1)=3.

Function values assigned to  coordinates in Figure \ref{fg_lattice_saw} 
are shown for completeness. 
They represent a special case of {\em index function} related to each coordinate.
Typically, they exhibit 1, 2, or 4 minima and have been designed
for performance testing of 
combinatorial algorithms~\cite{Lib-OPUS2-walk-2011-EV-Brglez}.
However, these values have no
particular significance in  Figure \ref{fg_lattice_saw} .
% it exhibits a single minimum and it turns into an extremely hard function when permuted
% Here, values
% are defined by a piecewise-linear {\em index function} that associates the function value
% with the coordinate position in the lexicographical order, offset by a `hidden integer' $i_H$
% to `hide' this relationship. For example, for $i_H = 0$ we would list the coordinate-value pairs
% as {\sf 00:0, 01:1, 02:2, 10:3 ...} etc. However, for $i_H = 4$ we list {\sf 11:0, 02:1, 20:2, 21:3 ...}
% etc, since $4 {(mod~4)} = 0$, $5 {(mod~4)} = 1$, etc. 

%% # fileName = ../xData/index/v-002-T.index 
%% #  command = funcT_index_gen 2 0 
%% #    created on Wed Jun 05 21:24:21 EDT 2013 
%% #    created coordinate pairs for coordType=T,  functionName=funcT_index, nVar=2, nFaces=9 
%% # .. lucky coord for min-value=0: 11 (luckyNum=4) 
%% idx	coord	index
%% 0	    00	5
%% 1 	01	6
%% 2 	02	7
%% 3	    10	8
%% 4	    11	0
%% 5    12	1
%% 6   	20	2
%% 7	    21	3
%% 8   	22	4  

In Figure \ref{fg_lattice_saw}-b on the left, the primitive cell is a cube, forming a unit cell as a stack of 3 cubes with 16 grid points.
On the right, we have a Hasse graph with 16 vertices with quarternary coordinate labels; this graph is not regular since
the degree of each vertex varies from 2, 3, to 4. 
In Figure \ref{fg_lattice_saw}-c on the left, the primitive cell is a cube, forming a unit cell as a large cube that contains 8 primitive cubes with 27 grid points.
On the right, we have a Hasse graph with 27 vertices with ternary coordinate labels; this graph is not regular since
the degree of each vertex varies from 3, 4, 5, to 6. 

In the context of this paper, it is important to also
study advances in self-avoiding walks being made in physics and elsewhere, for example on
the progression of improvements in the walk lengths
of self-avoiding walks on 2-, 3-, and 4-dimensional 
lattices~\cite{Lib-OPUS-walk-self-avoiding-2010-StatPhysics-Clisby-pivot-alg}.
%lattices~\cite{Lib-OPUS-walk-self-avoiding-1988-StatPhysics-Madras-pivot-alg,
%Lib-OPUS-walk-self-avoiding-2002-StatPhysics-Kennedy-pivot-alg,
%Lib-OPUS-walk-self-avoiding-2010-StatPhysics-Clisby-pivot-alg}.

\par\vspace*{1ex}\noindent
{\bf Protein folding examples.}
There are numerous articles that cover many more details about protein folding than we can present here,
from very technical~\cite{Lib-OPUS-bioinfo.fold-2009-InfSyst-Istrail-survey}
to tutorial~\cite{Lib-OPUS-bioinfo.fold-1998-AmSci-Hayes-Protein}. Our presentation
attempts to be generic and aims to make the problem accessible as a complex puzzle.

\begin{figure*}[]
\vspace*{-5ex}
\small
\begin{center}

\includegraphics[width=1.0\textwidth]{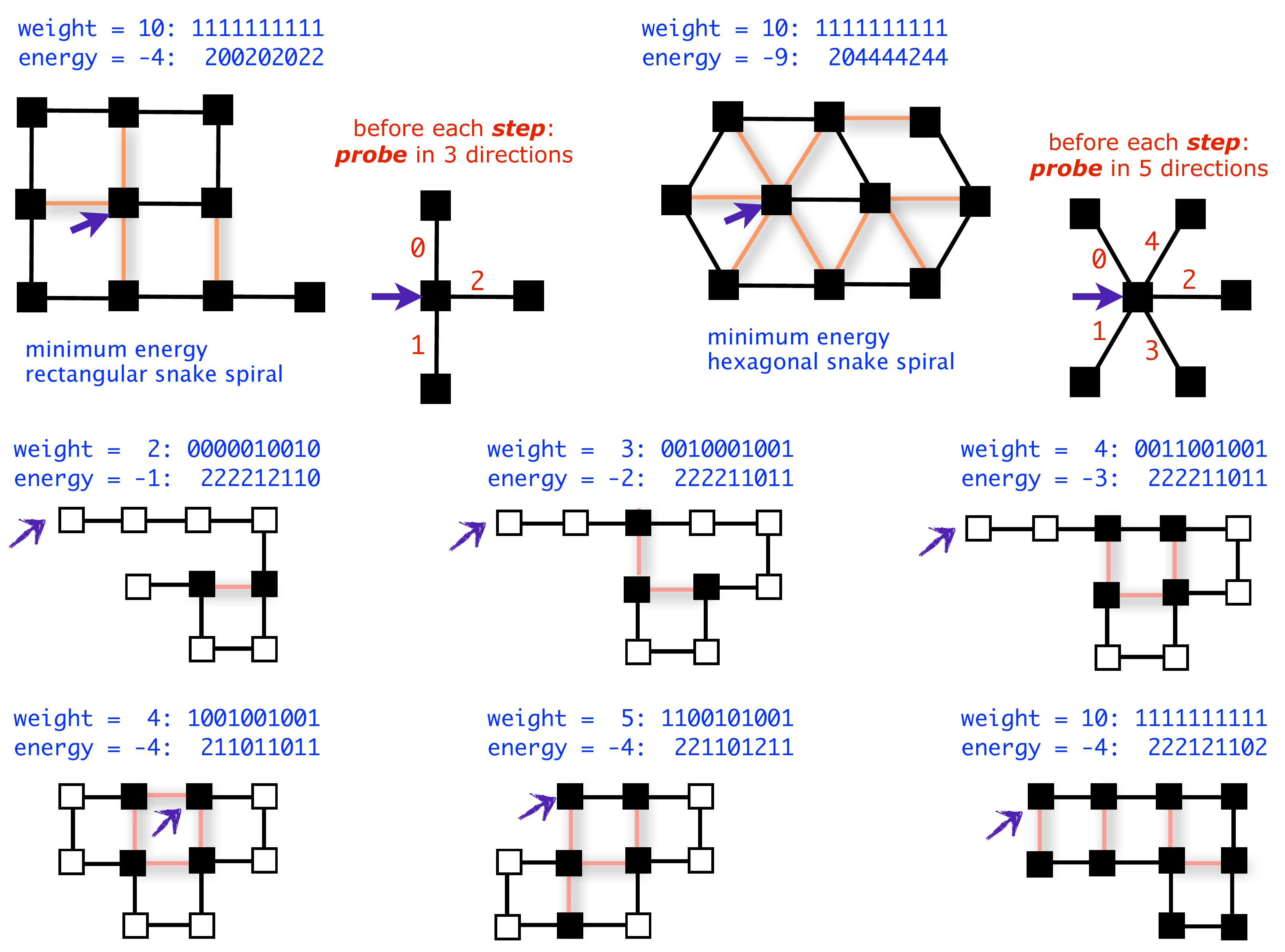}
\par\vspace*{1.5ex}
%\hspace*{-1.7em}  
\begin{tabular}{cccc | cccc | ccc} 

\multicolumn{11}{c}{
{\bf Six folding experiments under the weight and energy target shown, with 1000 seeds each}} \\[0.5ex] 
\hline
binary & energy & beyond & unique & \multicolumn{4}{c|}{walkLength} & \multicolumn{3}{c}{probesPerStep}  \\
weight & target &  target & solutions & median & mean  & stdev & max & median & mean  & stdev \\ 
\hline 
2 & -1 & 0 & 813 & 1000 & 843.6 & 695.5 & 2336 & 10.8 & 11.3 & 2.12 \\
3 & -2 & 0 & 511 & 39 & 338.0 & 580.7 & 2353 & 13 & 13.1 & 1.61 \\
4 & -3 & 95 & 204 & 26 & 104.1 & 290.7 & 2017 & 14.6 & 14.7 & 1.47 \\
4 & -4 & 0 & 2 & 797.5 & 1074.3 & 965.7 & 7433 & 13.9 & 14.0 & 0.82 \\
5 & -4 & 0 & 51 & 37.5 & 73.1 & 131.0 & 1755 & 15.8 & 16.0 & 1.27 \\
10 & -4 & 0 & 197 & 2 & 4.3 & 24.6 & 689 & 22 & 21.2 & 4.47 \\
\hline
\end{tabular}
\par\vspace*{0.75ex}
\includegraphics[width=1.0\textwidth]{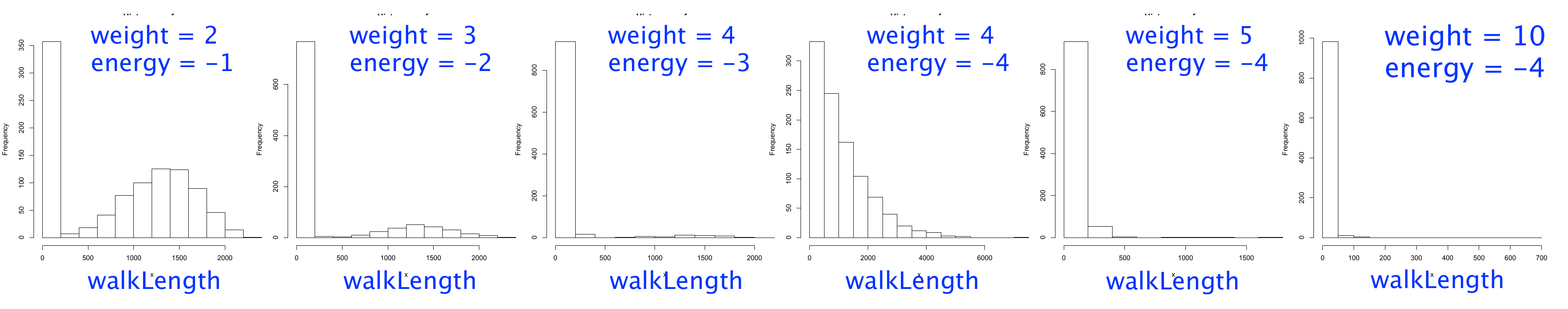}
%\vspace*{-1.5ex}
%\par
\caption{
Empirical observations about the HP model of protein foldings in 2D with a chain of size 10
and SAWs:
(1) lower bound on energy in rectangular grid is -4 ({\em probes} are made in 3 directions); 
(2) lower bound on energy in hexagonal grid is -9 ({\em probes} are made in 5 directions);
(3) instances of 6 folding conformations in rectangular grid, each for a distinct pair of weight and energy target,
define 6 classes of solutions; 
(4) walk length statistics and distributions, with a sample size of 1000 for each experiment.
In order to find the postulated energy targets, an exhaustive enumeration or a
Hamiltonian walk would visit, on the average, a total of 
$0.5*(2^{10}*3^9 - 1) = 10,077,696$ coordinates under the rectangular grid formulation (compare with the maximum of 7433 in the table) and a total of $0.5*(2^{10*}5^9 - 1) = 999,999,999$ coordinates under the hexagonal grid formulation.
Our hypothesis: compared to the results shown here, the hexagonal grid may exhibit an energy landscape where
SAWs will find energy minima in less steps on the average.
}
\label{fg_BT_conformations_intro}
\end{center}
\end{figure*}

Let us take $n$ beads in $k$ colors,
arrange them into a linear chain of length $n$ and register the position and the color of each bead, then allow the chain to fold
onto a predefined grid in a space of 2 or 3 dimensions. 
The most popular model is the 2-color HP (hydrophobic and polar, black and white, '1' and '0') model, where any pair of H-type beads that are adjacent on the grid after folding forms a bond. We measure the quality of the folding by counting the number of such bonds in a given arrangement, called a {\em conformation}. Once we subtract this number from 0, we call the value {\em energy} of the folding and the objective of any folding optimization algorithm is to minimize the value of this energy.
In Figure~\ref{fg_BT_conformations_intro} we display a number of chains, each of length 10, where the number of black beads varies from 2 to 10 and the energy from -1 to -4 (the maximum possible). An additional characteristic
of the chain is denoted as {\em weight}: it is simply the number of black beads in the chain. 

On a 2-dimensional {\em square lattice}, each step of a SAW has a choice of at most 3 adjacent points of the grid:
left, right, and forward, encoded as 0, 1, and 2. With the binary encoding of the colors and the ternary encoding of
the self-avoiding walk to control the folding, we encode the coordinate for the folding problem for a chain of black and white beads of length $n$ as a concatenation of $n$ binary and $n-1$ ternary coordinates, defining a hyperhedron with a total of  $(2^n)\times(3^{n - 1})$ face-centered vertices.
As an alternative, we may also choose a more efficient hexagonal lattice which, when the chain is folded, will produce more bonds  (the bees have done it!). In this case, there are 5 adjacent points on the grid; now the string of length $n$ is represented as a concatenation of $n$ binary and $n-1$ quinary coordinates, defining a hyperhedron with a total of  $(2^n)\times(5^{n - 1})$
face-centered vertices.

%\begin{figure*}[H] % this does work, i.e. [th!] (need to leave at least space of 3 lines at the bottom)
%\begin{figure}[H] % [] seems to work best ; this does NOT work [H]
\begin{table*}[t!]
\begin{center}
\caption{
Statistical experiments with SAWs to solve, under Plan A, Plan B and Plan C, the HP model of protein folding 
in 2D on a square lattice.
The  input parameters for each plan are:
Chain size = 10; Energy target = -4; either fixed binary coordinate  \texttt{coordB} of weight 4 (plan A); or fixed ternary coordinate \texttt{coordT} (plan B);
or both initialized randomly (plan C).
Experiments are performed with 1000 initial coordinates for each plan, and both the energy target of -4 and the binary weight target of 4 are reached under each plan, always returning only one binary solutions and two ternary solutions. Walk lengths under each plan differ significantly, with plan C representing the hardest instance of the folding protein problem. This is also the problem where SAW 
is the most effective compared to the (hypothetical) hamiltonian walk.
}
\label{tb_BT_conformations_plans}
\small
\vspace*{-3.1ex}
\begin{tabular}{|c@{~~~~}|r@{~~~~}|c@{~~~~}c@{~~~~}c@{~~~~}c@{~~~~}c@{~~~~}|}
\multicolumn{7}{c}{
{\bf  }} \\[0.5ex]  
\hline
{\textbf{Plan A}}  &  & median & mean & stdev & min & max \\
\hline
Given {\bf coordB}      & cntProbe & 244 & 330.6 & 302.4 & 10 & 1994\\
with weight = 4,   & walkLength & 25 & 33.7 & 31.4 & 1 & 205\\
{\bf FIND coordT}        & probesPerStep & 10 & 10.4 & 1.2 & 7.9 & 16\\
\texttt{coordB = 1001001001} & \multicolumn{6}{c|}{}\\
\texttt{coordT = ~200100100} & \multicolumn{6}{c|}{The average walkLength to reach one of the two \texttt{coordT}}\\
\texttt{coordT = ~211011011} & \multicolumn{6}{c|}{under Hamiltonian walk = $0.25(3^{9} - 1) = 4921$}\\
\end{tabular}

\vspace*{-1.1ex}
\begin{tabular}{|c@{~~~~}|r@{~~~~}|c@{~~~~}c@{~~~~}c@{~~~~}c@{~~~~}c@{~~~~}|}
\multicolumn{7}{c}{
{\bf  }} \\[0.5ex]  
\hline
{\textbf{Plan B}}  &  & median & mean & stdev & min & max \\
\hline
Given {\bf coordT}, & cntProbe & 26 & 34.0 & 15.6 & 1 & 127\\
{\bf FIND coordB}& walkLength & 4 & 5.2 & 2.4 & 0 & 20\\
with weight = 4 & probesPerStep & 6.5 & 6.5 & 0.49 & 1 & 8.5\\
\texttt{coordT = ~211011011} & \multicolumn{6}{c|}{The average walkLength to reach the single \texttt{coordB}}\\
\texttt{coordB = 1001001001} & \multicolumn{6}{c|}{under Hamiltonian walk = $0.5(2^{10} - 1) = 511$}\\
\end{tabular}

\vspace*{-1.1ex}
\begin{tabular}{|c@{~~~~}|r@{~~~~}|c@{~~~~}c@{~~~~}c@{~~~~}c@{~~~~}c@{~~~~}|}
\multicolumn{7}{c}{
{\bf  }} \\[0.5ex]  
\hline
{\textbf{Plan C}}  &  & median & mean & stdev & min & max \\
\hline
{\bf FIND coordB}, & cntProbe & 10564.5 & 14187.8 & 12787.2 & 35 & 81156\\
with weight = 4 & walkLength & 752.5 & 1027.1 & 936.3 & 2 & 5936\\
\& {\bf FIND coordT} & probesPerStep & 13.9 & 14.0 & 0.7 & 12.0 & 19.6\\
\texttt{coordB = 1001001001} & \multicolumn{6}{c|}{}\\
\texttt{coordT = ~200100100} & \multicolumn{6}{c|}{The average walkLength to reach one of the two \texttt{coordT}}\\
\texttt{coordT = ~211011011} & \multicolumn{6}{c|}{under Hamiltonian walk = $0.25*(2^{10}*3^9 -1) = 5038848$}\\            
\end{tabular}
\end{center}
\vspace*{-4.0ex}
\end{table*}

In this paper, we experiment with foldings on a 2-dimensional square lattice. We have grouped our  experiments under three plans: 
\begin{description}
\item [Plan A]
Strech a chain of length $n$ with weight $w$ and
assign the $w$ black beads into fixed positions.
Represent this chain as a binary coordinate of length $n$ and weight $w$. 
Search for folding of this chain on a square lattice in 2D that will minimize its energy.
Represent the solution as a ternary coordinate of length $n-1$.

\item [Plan B]
Fold a chain of length $n$ with weight $w=n$ into
a preferred conformation. Typically, the preferred conformation is the one
where the energy, with all beads being black, is the global minimum.
Two such conformations, with all beads being black and the energy of -4, are shown in 
Figure~\ref{fg_BT_conformations_intro}. 
Now, represent this conformation as ternary coordinate of length $n$.
Search for a binary coordinate of weight $w < n$ that 
either retains the minimum energy under all beads being black or is
as close as possible to this value.

\item [Plan C]
Select the length of the chain $n$, its weight $w <= n$, and the
target energy value that can be satisfied with at least one feasible
folding conformation.
Assign a random binary string of length $n$ and weight $w$ as the
initial binary coordinate. Assign a random ternary ternary string
of length $n -1$ as the initial ternary coordinate.
Chances are that some initial ternary coordinates do not represent
a feasible folding -- this is not an issue since
in our formulation, the search escapes the unfeasible regions effectively.
The search now proceeds by probing  simultaneously
segments of each concatenated coordinate: the binary segment and the ternary segment
before returning a feasible solution with the given weight and the energy target value.

\end{description}

Plan A represents the traditional formulation of the folding problem
and many experimental results are reported under this plan.
Plan B is also known as the {\em inverse folding problem} formulation
and  experimental results  are also reported under this plan.
A number of experiments that rely on {\em exhaustive enumeration} have similarities with Plan C.
However, 
we are not aware of any publication of experimental results as described under Plan C
in this paper; if brought to our attention, we shall report on them
in our future publication.

%\begin{figure*}[H] % this does work, i.e. [th!] (need to leave at least space of 3 lines at the bottom)
\begin{figure*}[t!]
%\small
\begin{center}
\vspace*{-2.0ex}
\hspace*{-1.8em}
\includegraphics[width=1.01\textwidth]{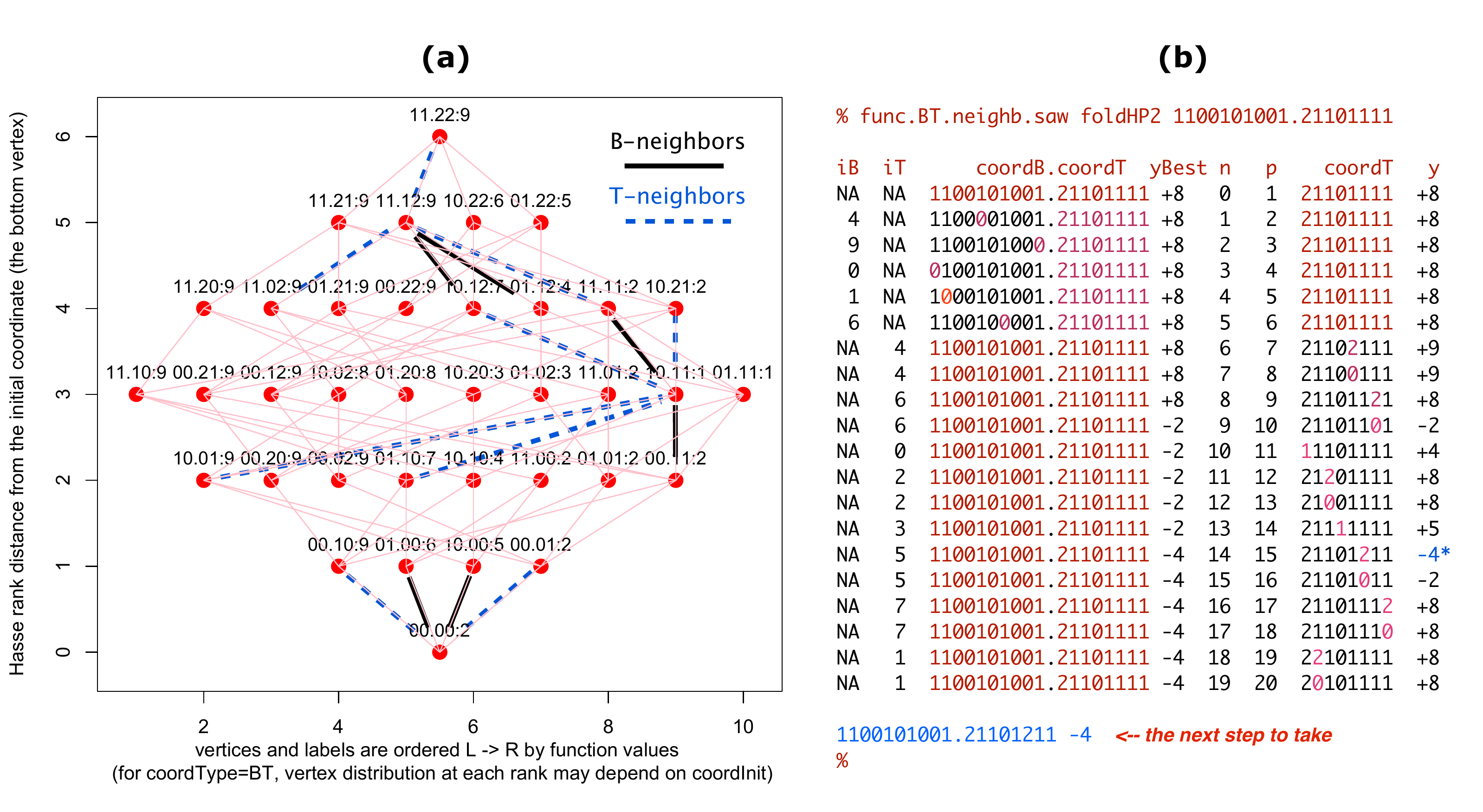}
\par\vspace*{-3ex}

\begin{tabular}{p{0.48\textwidth}p{-0.00\textwidth}p{0.45\textwidth}}
\begin{tabular}{p{0.48\textwidth}}

{\footnotesize
\begin{center}
\begin{minipage}[]{0.48\textwidth} 
Introduced in Figure~\ref{fg_BT_dice}, labels in this graph are a
a concatenation of binary coordinates and ternary coordinates.
Each binary coordinate always has a neighborhood of 2 (dotted edges) while the neighborhood of ternary coordinate can vary from 2 to 4 (solid edges). For example, the coordinate 00.00 has 2 binary and two ternary neighbors;
the coordinate 10.11 has 2 binary and 4 ternary neighbors.
\end{minipage}
\end{center}
}
\end{tabular}
&
\begin{tabular}{p{-0.00\textwidth}}
\par
\end{tabular}
&
\begin{tabular}{p{0.45\textwidth}}

{\footnotesize
\begin{center}
\begin{minipage}[]{0.45\textwidth} 

Indices $iB$ and $iT$ that address values in binary and ternary
coordinates are always randomly permuted in order to prevent
biasing the order of choices for best function value {\em yBest}.
Function values $y > 0$ represent not only unfeasible
conformations but also the relative level of unfeasibility.
The counters $n$ and $p$ report the size of the neighborhood
and the number probes to find each value of~$y$.
\end{minipage}
\end{center}
}
\end{tabular}
%}
\end{tabular}
\par\vspace*{-0.5ex}
\caption{
An example of neighborhood calculation from an initial pivot 
coordinate (1100101001.21101111) that leads, in a single step, to an optimal 
conformation depicted in Figure \ref{fg_BT_conformations_intro}.
}\label{fg_BT_neighborhood_calc}
\end{center}
\vspace*{-4.0ex}
\end{figure*}

The summary of statistical experiments in Figure~\ref{fg_BT_conformations_intro}
reveals a number of interesting properties.
All have been performed under Plan C, with 1000 randomly selected initial configurations
for each of the six weight and energy input pairs:
\begin{itemize}
\item
As the tabulated binary weight increases and the energy target value decreases, 
the number of unique solution decrease from 813 (out of 1000 trials) to 2 (for weight = 4 and energy = -4),
but then increase to 197 (for weight = 10 and energy = 4).
The walkLength statistics varies significantly for each case -- as does the distribution, which is bimodal, heavy-tailed, and clearly geometric for the case of only 2 unique solutions with weight = 4 and energy = -4.
\item
The beneficial side-effect our testing strategy is revealed for the case of weight = 4 and energy target = -3.
Out of 1000 trials, there are 95 conformations with energy = -4, i.e. the returned solution is better than the target solution of -3. These solutions are in the class of 'rare solutions' where only two unique solutions have been reported after 1000 trials for weight = 4 and energy = -4.
We shall take advantage of this strategy also when performing experiments on longer chains which are summarized in
Section~\ref{sec_experiments}. 
%\item
%An exhaustive experiment would require an evaluation of
%$(2^{10}*3^9 - 1) = 20,155,391$ coordinates. The worst case averages for {\em walkLength} and {\em probesPerStep}
%in Figure ~\ref{fg_BT_conformations_intro} are 1074.3 and 14.0, thus the average number of coordinate evaluations %for each trial is 1074.3*14.0 = 15,040 and 15,040,000 for 1000 trials. The effectiveness of SAWs will be %demonstrated more dramatically on chains that are longer than 10 in Section~\ref{sec_experiments} where the %exhaustive evaluations are not even an option.
\end{itemize}
We complete this subsection with the summary of results under Plan A, Plan B, and Plan C, shown in Table~\ref{tb_BT_conformations_plans}. Notably, the search with SAW under the plan B (the inverse folding formulation) is significantly easier than the search under Plan A (the traditional folding formulation). However, the search with SAW under Plan C requires significantly more steps than the Plan A and Plan B combined. The experiment under Plan C in Table~\ref{tb_BT_conformations_plans} is a replication, under a different initial seed, of the experiment in 
Figure~\ref{fg_BT_conformations_intro} in the row  weight = 4, energy = -4.

\par\vspace*{1ex}\noindent
{\bf Global stochastic search under SAW.}
We now briefly formalize coordinate neighborhoods, walks, and self-avoiding walks, 
concluding with a concise pseudo code that is the basis for our
prototype solver on stochastic search under SAW.

\par\vspace*{0.5ex}\noindent
{\em Coordinate neighborhood.}
Formally, a neighborhood of a coordinate  ${\underline \varsigma}_j$  is a set of coordinates
\begin{equation}
  {\cal N} ({\underline \varsigma}_j) = \{~ {\underline \varsigma}_j^i ~|~ d({\underline \varsigma}_j, {\underline \varsigma}_j^i ) = 1, 
  ~~~i = 1, 2, \dots , L_j ~\}
\label{eq_neighborhood}
\end{equation}
where $d({\underline \varsigma}_j, {\underline \varsigma}_j^i )$ is the rank distance between coordinates.
The coordinate ${\underline \varsigma}_j$ is also called a {\em pivot coordinate}, has $L_j$ 
neighbors, each a distance of 1 from the pivot coordinate. 
In Figure~\ref{fg_BT_neighborhood_calc}-a we illustrate a Hasse graph that highlights three neighborhoods.
Coordinates in this graph are a concatenation of binary coordinates of length 2 and ternary coordinates of length 2.
Each binary coordinate always has a neighborhood of 2 (dotted edges) while the neighborhood of ternary coordinate can vary from 2 to 4 (solid edges). For example, the coordinate 00.00 has 2 binary and two ternary neighbors;
the coordinate 10.11 has 2 binary and 4 ternary neighbors.

In Figure~\ref{fg_BT_neighborhood_calc}-b we illustrate the dynamics of neighborhood evaluations for an
instance shown in Figure~\ref{fg_BT_conformations_intro}. We could not possibly have drawn a Hasse graph for this instance, however the principle of binary and ternary neighborhoods illustrated
in Figure~\ref{fg_BT_neighborhood_calc}-a are the same.

What we can show is the trace of the entire 
neighborhood evaluation
that takes place: starting with the pivot coordinate 1100101001.21101111, the best coordinate of the next
pivot in this neighborhood is 1100101001.21101211 since it has the best energy conformation of -4. The trace also shows values of the objective function for various conformations -- all values that are > 0 represent 
conformations that {\em would} lead to a conflict during folding; penalties values are assigned at different levels: +8 (initial pivot), +9, +8,
+4, etc. Not all binary coordinates have been evaluated due to an input requirement that
weight <= 5. 
%During the walk, this neighborhood would `shrink' also if there are any pivots in the %neighborhood.
The situation where a pivot would get trapped by adjacent pivots and the  neighborhood
would become empty did not yet arise.

\begin{figure*}[!t]
%\small
\begin{center}
\vspace*{-2ex}
%{\bf (a)} \hspace{40ex} {\bf (b)}
\begin{tabular}{p{0.45\textwidth}p{0.52\textwidth}}
\hspace*{-2.3em}
\mbox{{
\begin{tabular}{p{0.45\textwidth}}
\begin{algorithmic}[1]

\State $s    \gets 1901 $                                      \Comment{initial seed}
\State ${\underline \xi}_0  \gets coordInit(s)$                \Comment{initial coordinate}
\State $\Theta({\underline \xi}_0)   \gets {\underline \xi}_0$ \Comment{initial value}
\State ${\underline \xi^*}         \gets {\underline \xi}_0$   \Comment{initial best coordinate}
\State $\Theta({\underline \xi^*}) \gets {\underline \xi^*}$   \Comment{initial best value}
\State $\omega \gets 0$                                        \Comment{initial walk length}
\State $\Theta^{ub}_L\gets 0$                                  \Comment{initial upper bound}
\State $\tau \gets 1$                                          \Comment{initial cntProbe}
\State $\tau_{lmt} \gets 2^{24}$                                 \Comment{cntProbe limit value}
\State $isCens \gets 0$                                        \Comment{initialize as uncensored}
\If {$\Theta({\underline \xi^*})  \le \Theta^{ub}_L$}
    \State $Table \gets (s, {\underline \xi^*}, \Theta({\underline \xi^*}), \tau, \omega, isCens)$
    ; \textbf{return}
     
\EndIf
\While {$\Theta({\underline \xi^*})  > \Theta^{ub}_L$}
    \If {$\tau == \tau_{lmt}$}  
        \State $isCens \gets 1$ 
        ; \textbf{break}
    \Else
       \State $\omega=\omega+1$                               \Comment{{\bf a new step!}}
       \State $temp \gets {\tt coordUpdate}({\underline \xi}_{\omega - 1} ,\tau )$ 
       \State ${\underline \xi}_{\omega}:\Theta({\underline \xi}_{\omega}):\tau  \gets temp$
       %\State $\Theta({\underline \xi}) \gets {\underline \xi}$
       \If {$\Theta({\underline \xi}_{\omega}) \le \Theta({\underline \xi^*})$} 
           \State ${\underline \xi^*}         \gets {\underline \xi_{\omega}}$
           \State $\Theta({\underline \xi^*}) \gets {\underline \xi^*}$
       \EndIf
    \EndIf
\EndWhile

\If {$isCens == 1$}
   \State $Table \gets (s, {\underline \xi^*}, \Theta({\underline \xi^*}), \tau, \omega, isCens)$
\Else
    \If {$\Theta({\underline \xi^*})  == \Theta^{ub}_L$}
        % x* <- x ; tau* <- tau
        \State $Table \gets (s, {\underline \xi^*}, \Theta({\underline \xi^*}), \tau, \omega, isCens)$
    \Else     
        % x** <- x ; tau** <- tau
        \State $\Theta^{ub}_L \gets \Theta({\underline \xi^*})$ \Comment{{\bf Better upper bound!}}
        \State $Table \gets (s, {\underline \xi^*}, \Theta({\underline \xi^*}), \tau, \omega, isCens)$
    \EndIf
\EndIf
    
\end{algorithmic}
\end{tabular}
}}
&
\begin{tabular}{p{0.52\textwidth}}
\begin{center}
\hspace*{-2.5em}
\begin{minipage}[]{0.52\textwidth}
{\small
The procedure {\tt coordUpdate.saw} takes the pivot coordinate, 
the probe counter and the walk list. In Step 6, it computes,
in random order, the neighborhood 
${\cal N} ({\underline \varsigma}_{\omega -1})$
of all adjacent coordinates.
The order randomization ensures that all coordinates get an equal chance of
selection; without it, some paths in the Hasse graph may never be taken,
thereby inducing a bias in the average walk length.
The Step 7 eliminates all adjacent coordinates that may have been 
used as pivots already and returns a neighborhood subset
${\cal N}_r ({\underline \varsigma}_{\omega -1})$.
If the neighborhood subset is not empty, the procedure 
{\tt bestNeighbor} in Step 9 probes all coordinates in the subset and returns
the new pivot as the {\sf coordinate:value pair} with the {\em `best value'}, along with the
incremented value of $\tau$, updates 
the walk list to $Walk_{\omega}$ in Step 10, and exits on Step 18.
An empty neighborhood implies that the SAW is {\em trapped}, 
i.e. the selection of the pivot for the next step is blocked by adjacent coordinates 
that are already pivots. Subsequently, a new walk segment is initialized with a random coordinate in Step 15. 
The procedure exits  with the expected parameter values on Step 16.
}
\end{minipage}
\end{center}

\mbox{{
\hspace*{-2.2em}
\begin{tabular}{p{0.52\textwidth}}
\begin{algorithmic}[1]
%% http://tex.stackexchange.com/questions/74849/increase-spacing-in-an-algorithm	
%% use \caption{\strut My algo} Ğ Herbert Oct 1 '12 at 10:36
%% \STATE \strut instruction 1 The command \strut
%% is the height of the character "(" but without a width. It can be used to get some more space, if needed Ğ Herbert Oct 1 '12 at 10:49
\par\vspace*{-3ex}
\State \strut $\omega \gets \omega + 1$
\State \strut $Walk_{\omega -1} \gets \{{\underline \varsigma}_0,~{\underline \varsigma}_1,~{\underline \varsigma}_2,~\dots ,~{\underline \varsigma}_{\omega -1}\}$

\Procedure{coordUpdate.saw}{${\underline \varsigma}_{\omega -1},\tau,Walk_{\omega -1}}$

\State \strut $\mathbb{Z} \gets i = 1,2, \ldots ,L $ 
\State \strut $\mathbb{Z}_p \gets permute(\mathbb{Z}) $ 
\State \strut ${\cal N} ({\underline \varsigma}_{\omega -1}) \gets  \{{\underline \varsigma}_{\omega -1}^i ~|~ d({\underline \varsigma}_{\omega -1}, {\underline \varsigma}_{\omega -1}^i ) = 1, ~i \in \mathbb{Z}_p\}$ 
\State \strut ${\cal N}_{r} ({\underline \varsigma}_{\omega -1}) \gets  \{ {\cal N} ({\underline \varsigma}_{\omega -1}) ~|~
                      {\underline \varsigma}_{\omega -1}^i \not\in  Walk_{\omega -1}\}$
   \If {\strut  ${\cal N}_{r} \not= \emptyset$}  
		\State \strut $\!\!\!\!{\underline \varsigma}_{\omega}\!\!:\!\Theta({\underline \varsigma}_{\omega})\!:\!\tau  
		\gets  bestNeighbor({\cal N}_{r}, {\underline \varsigma}_{\omega -1}, \tau_{\omega -1})$
		\State \strut $\!\!\!\!Walk_{\omega} \gets \{{\underline \varsigma}_0,~{\underline \varsigma}_1,~{\underline \varsigma}_2,~\dots ,~{\underline \varsigma}_{\omega -1},~{\underline \varsigma}_{\omega}\}$
    \Else \Comment{\textbf{trapped pivot, restart}}
		\State \strut $\!\!\!\!s \gets randomSeed()$ 
		\State \strut $\!\!\!\!{\underline \varsigma}_0          \gets coordInit(s)$     \Comment{new initial coord.}
		\State \strut $\!\!\!\!\Theta({\underline \varsigma}_0)   \gets {\underline \varsigma}_0$ \Comment{new initial value}
		\State \strut $\!\!\!\!Walk_{0} \gets \{ {\underline \varsigma}_0 \}$               \Comment{new walk segm.}
		\State \strut \!\!\!\!\textbf{return} ${\underline \varsigma}_{0}\!\!:\!\Theta({\underline \varsigma}_{0})\!:\!\tau\!:\!Walk_{0}$
    \EndIf
\State \strut \textbf{return} ${\underline \varsigma}_{\omega}\!\!:\!\Theta({\underline \varsigma}_{\omega})\!:\!\tau\!:\!Walk_{\omega}$ 

\EndProcedure 
\end{algorithmic}
\end{tabular}
}}
\end{tabular}
\end{tabular}
\vspace*{-3ex}
\caption{The walk as a part of the {\bf global stochastic search} process: the walk stops 
(1) upon reaching the best upper bound, returning a new or already known solution coordinate, or
(2) upon finding a new best upper bound, returning a new best solution coordinate, or 
(3) upon exceeding the allocated time of counter limit, returning a new or already known censored solution coordinate and a value above the upper bound. 
The procedure that controls the performance of the walk, here a {\em self-avoiding walk}, is named \coordUpdate.
%Result shown in the
%three tables summarize the case when \coordUpdate~ implements a contiguous {\bf random walk solver} on instance of 
%the \labs\ problem of size $L=13$.
%The instance itself is described under Figure \ref{fg_walks_restarts_vs_saw}.
}
\label{fg_global_search}
\vspace*{-2ex}
\end{center}
\end{figure*}

\par\vspace*{0.5ex}\noindent
{\em Contiguous walks and SAWs.}
%We also introduce the concept of a walk. % a contiguous walk in particular.
Let the coordinate ${\underline \varsigma}_0$ be the initial coordinate from which the walk takes the first step. Then the sequence  
\begin{equation}
  \{{\underline \varsigma}_0, {\underline \varsigma}_1, {\underline \varsigma}_2, \dots , {\underline \varsigma}_j, \dots , {\underline \varsigma}_\omega\}
\label{eq_coord_pivots}
\end{equation}
is called a {\em walk list} or a {\em walk} of length $\omega$, the coordinates ${\underline \varsigma}_j$ are denoted as {\em pivot coordinates} and $\Theta({\underline \varsigma}_j)$ are denoted as {\em pivot values}.
Given an instance of size $L$ and its best upper bound $\Theta^{ub}_L$,
we say that the walk {\em reaches} its target value (and stops) when
$\Theta({\underline \varsigma}_\omega) = \Theta^{ub}_L$.

We say that the walk is contiguous if the rank distance between adjacent pivots is 1; i.e. we find
\[ d({\underline \varsigma}_j, {\underline \varsigma}_{j-1}) = 1, ~~~j = 1, 2, ..., \omega \] 

We say that the walk is {\em self-avoiding} 
if all pivots in (\ref{eq_coord_pivots}) are unique.
%$\rho$ 
We say that the walk is composed of two or more {\em walk segments} if the initial pivot of each walk segment has been induced by a well-defined heuristic such as {\em random restarts}.
Walk segments can be of different lengths and
if viewed independently of other walks,
may be self-avoiding or not. A walk composed of two or more self-avoiding walk segments may no longer be a self-avoiding walk, since some of the pivots may overlap and also form cycles.
%If the walk is composed of three segments, two segments have been induced by two random restarts. 

\par\vspace*{0.5ex}\noindent
{\em Global stochastic search under SAW.}
The pseudo-code in Figure~\ref{fg_global_search} formalizes
the global search algorithm that relies on SAW as its search engine.
The code forms the basis for the prototype solver not only for
the porting folding instances experiments in this paper but also
for a number of other problem instance as outlined in Section~\ref{sec_introduction}. 

%\lipsum[3] % Dummy text

\section{Summary of Experiments}
\label{sec_experiments}
\par
\begin{figure*}[t!]
%\small
\begin{center}
\vspace*{-3ex}
%\hspace*{-2.5ex}
\includegraphics[width=1.00\textwidth]{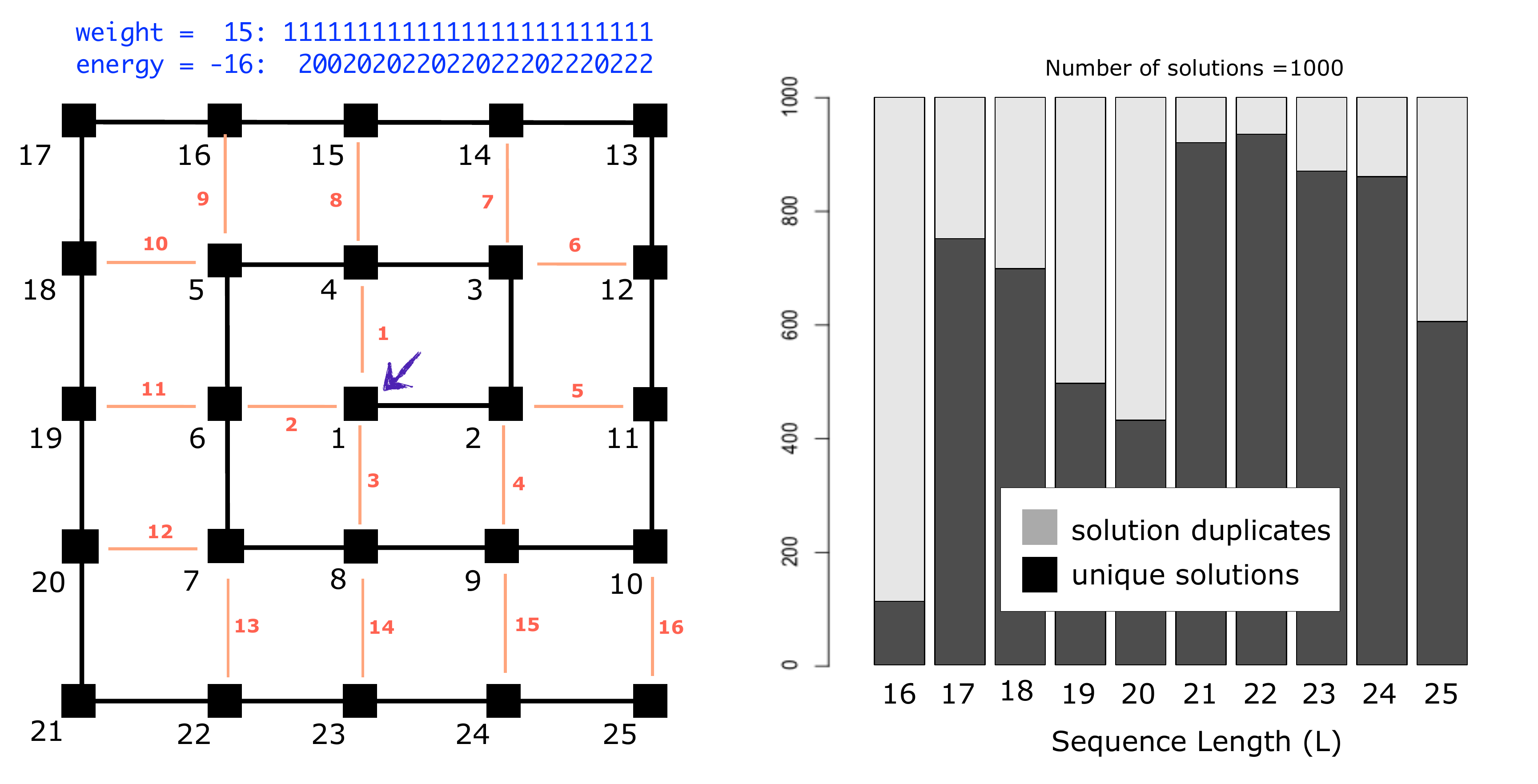}
\\[-3.5ex]
%\hspace*{-6.5ex}
\includegraphics[width=1.00\textwidth]{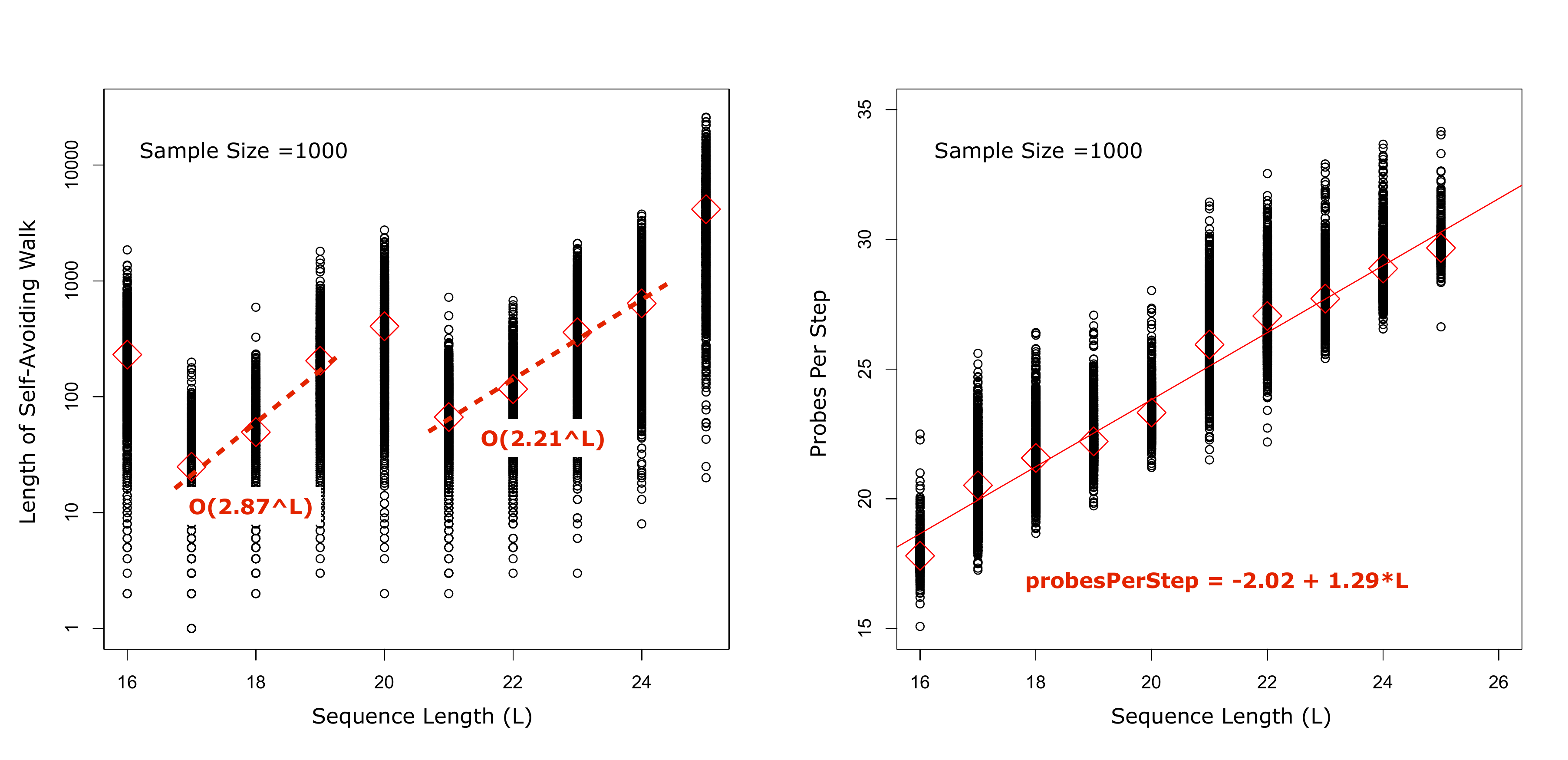}

\vspace*{-3.0ex}
\caption{
The spiral chain can be considered as a simpler case of the perfect 
HP problem posited for a 3-D cube 
in~\cite{Lib-OPUS-bioinfo.fold-1998-MIT-Leighton-perfect_HP_string}.
In our context, we ask: how many conformations can be found with the same energy and how
hard is it to find them? Some of the answers can be gleaned from the 
asymptotic performance of the SAW solver in terms of the required 
walkLength to reach the minimum energy targets of (-9, -10, ..., -16)
for chain lengths for L = (15, 17, ..., 25). It appears that there are clear different
walkLength performance regimes when solving with L = (16, 20, 25) when compared to
solving with other values of L.
}
\label{fg_snake_spiral}
\end{center}
\vspace*{-4.0ex}
\end{figure*}

\begin{figure*}[]
%\small
\begin{center}
\vspace*{-2ex}
\hspace*{-2.5ex}
\includegraphics[width=1.02\textwidth]{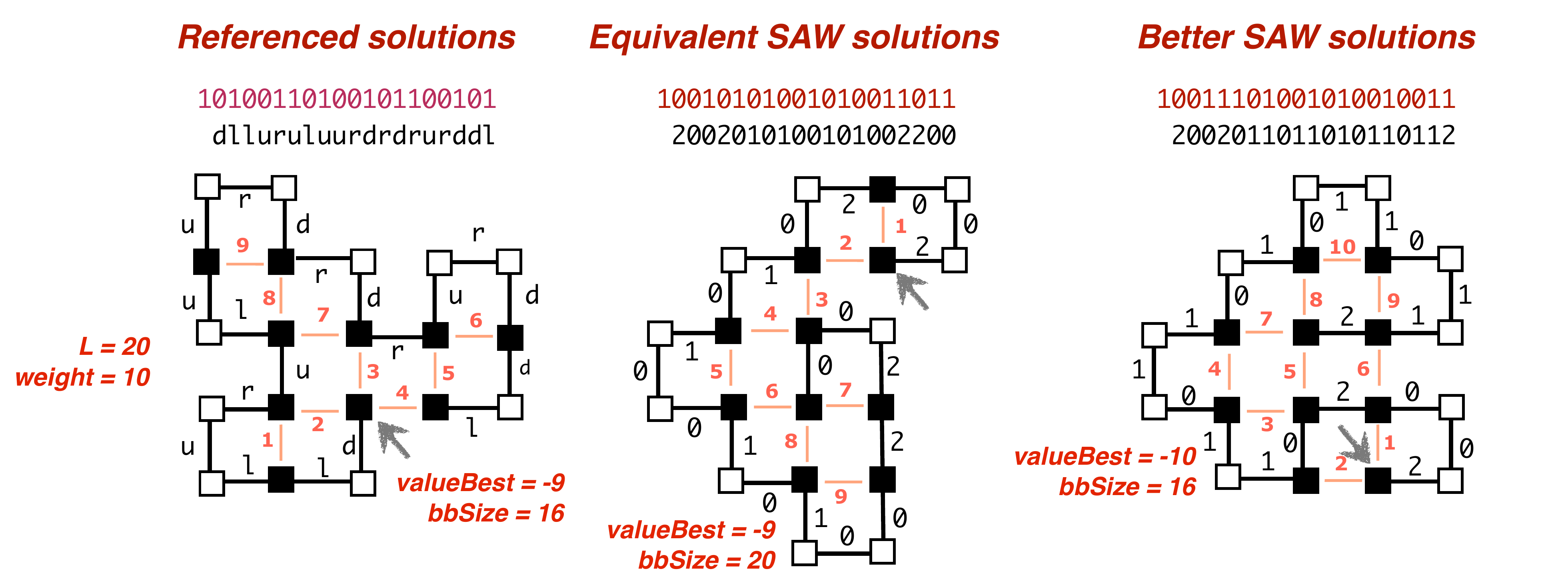}
\\[0.5ex]
\hspace*{-2.5ex}
\includegraphics[width=1.02\textwidth]{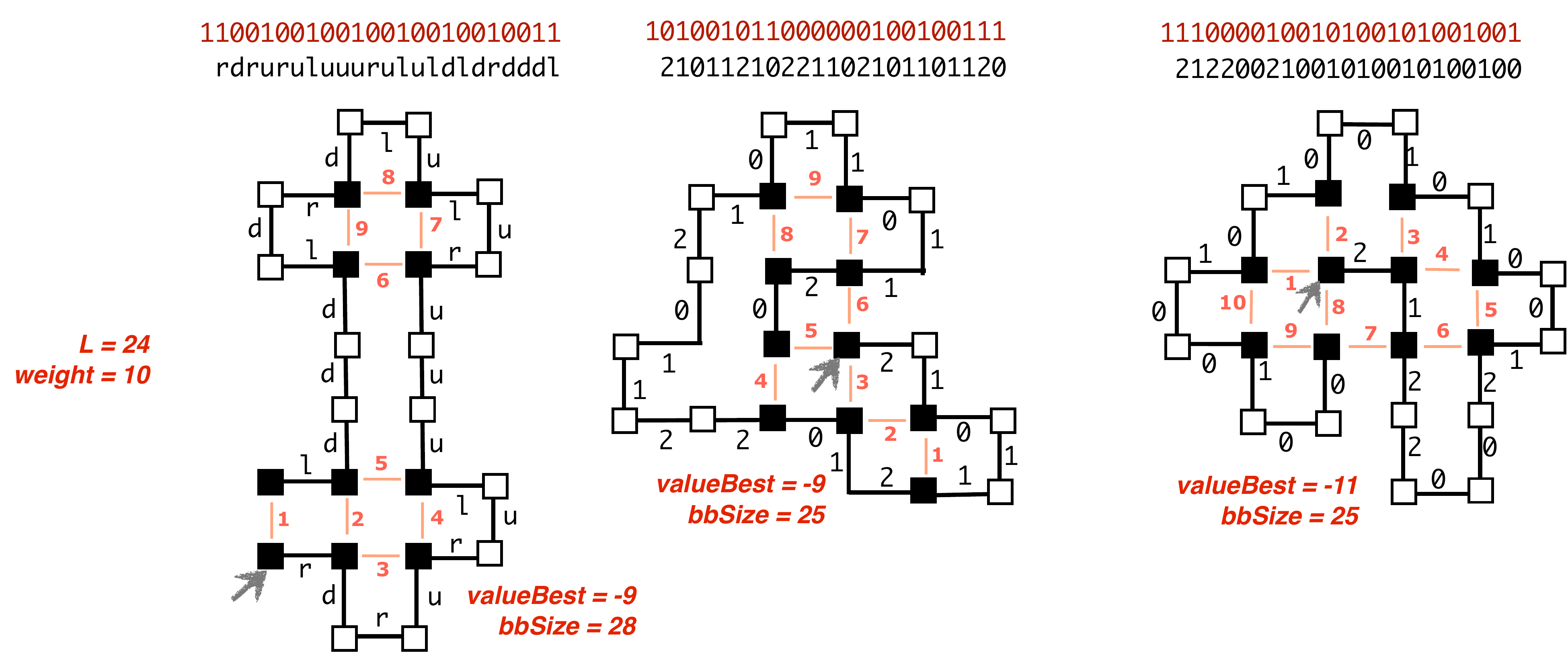}
\\[0.5ex]
\hspace*{-2.5ex}
\includegraphics[width=1.02\textwidth]{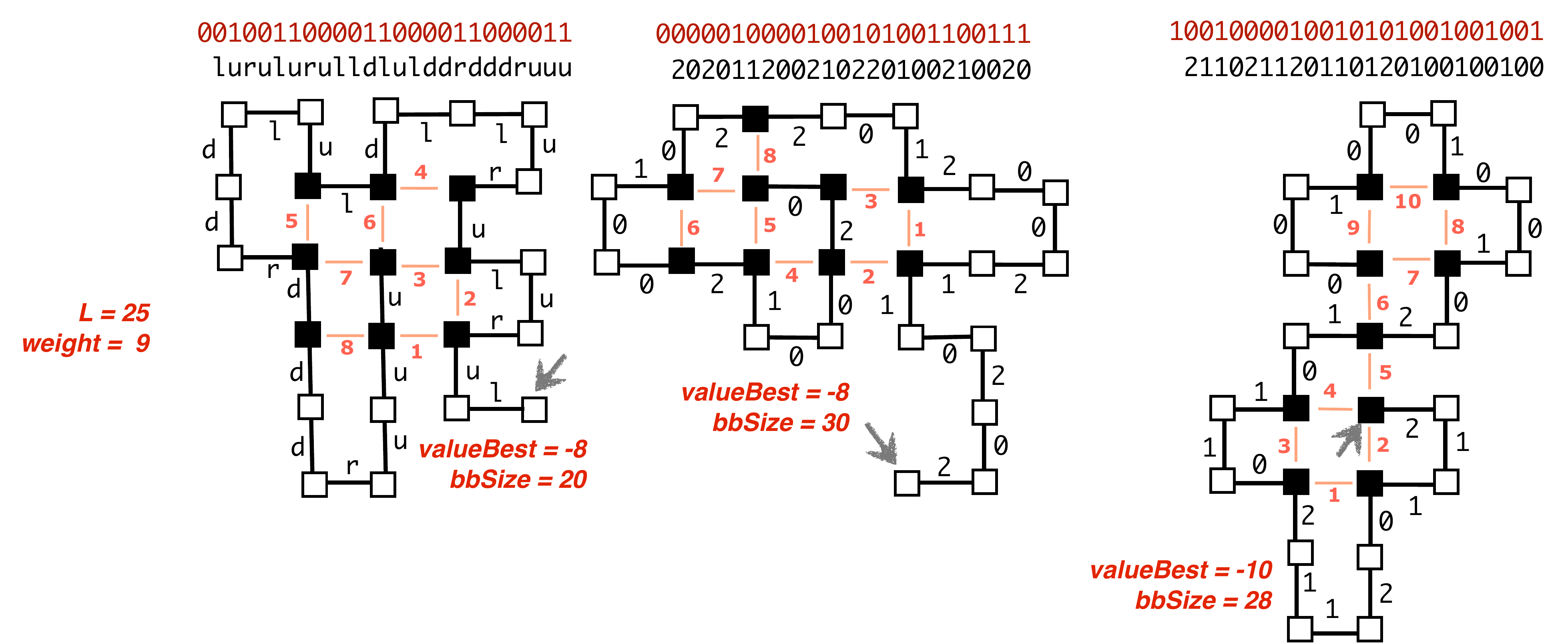}
\caption{Comparisons of protein folding conformations for chain lengths of 20, 24, and 25.
instance conformation in the column `Referenced solutions' are from
from~\cite{Lib-OPUS-protein-folding-1993-JMB-Unger} 
and~\cite{Lib-OPUS-protein-folding-2005-GECCO-Bui}
and have been reported under what we call Plan A in this paper.
Instances under the column `Equivalent SAW solutions' are alternative
foldings obtained by our SAW solver under Plan C
and the same energy targets as shown for `Referenced solutions'.
Instances under the column `Better SAW solutions' are alternative
foldings obtained by our SAW solver under Plan C
and {\em better} energy targets:
-10 vs -9 for L=20; -11 vs -9 for L=24, and -10 vs -8 for L=25.
}
\label{fg_BT_conformations}
\vspace*{-4ex}
\end{center}
\end{figure*}

\vspace*{-0.95ex}\noindent
%\lipsum[1] \par\lipsum[2]  \par\lipsum[3] % Dummy text
Experimental results, summarized in 
Figure~\ref{fg_snake_spiral}, Figure~\ref{fg_BT_conformations}, and Table~\ref{tb_BT_conformations},
can only be descried briefly.

The motivation for the experiment in Figure~\ref{fg_snake_spiral} came
from~\cite{Lib-OPUS-bioinfo.fold-1998-MIT-Leighton-perfect_HP_string}:
\begin{center}
\begin{minipage}[]{0.46\textwidth}
{\small
{\em Instance:} An integer $n$ and a finite sequence of $S$ over the alphabet $\{H, P\}$ which contains $n^3$ $H$'s.
\\
{\em Question:} Is there a fold of $S$ in $\mathbb{Z}^3$ for which H's are perfectly
packed into an $n \times n \times n$ cube?
\par
~~~This problem has been proven to be NP-hard and the more general problem as NP-complete.
}
\end{minipage}
\end{center}
The spiral chain in Figure~\ref{fg_snake_spiral} can be considered as a simpler case of the perfect 
HP problem posited for a 3-D cube 
in~\cite{Lib-OPUS-bioinfo.fold-1998-MIT-Leighton-perfect_HP_string}.
In our context, we ask: how many conformations can be found with the same energy and how
hard is it to find them? The answers that we display
in Figure~\ref{fg_snake_spiral} are somewhat surprising and will be analyzed in more depth later.

%\begin{figure*}[H] % this does work, i.e. [th!] (need to leave at least space of 3 lines at the bottom)
%Also refer
%to~\cite{Lib-OPUS-bionfo.fold-2008-Springer-Boekenhauer-triangle} and
%to~\cite{Lib-OPUS-bioinfo.fold-2009-InfSyst-Istrail-survey}
%for more details ....
%As to~\cite{Lib-OPUS-bionfo.fold-2008-Springer-Boekenhauer-triangle}:
%\\
%-- our results are based on square grin can only be compared to results such
%as xxx, xxx, xxx, where were replicated current results and also exceeded the quality of 
%solutions ... Runtimes may be compared and scaled to larger instances later, when our implementation
%in C is completed.

%-- Our work with SAWs 
%on the notoriously hard labs problem \cite{labs-refs} 
%is being compared to memetic//tabu approach and 
%results with SAWs deliver results of better quality under the
%same time constraint (on the same platform).

%-- It will be interesting to compare SAWs based on hexagonal grid to
%results reported in
%in~\cite{Lib-OPUS-bionfo.fold-2008-Springer-Boekenhauer-triangle}.

Experiments performed on well-known
instance classes under Plan A and Plan C
are summarized in
Figure~\ref{fg_BT_conformations} and Table~\ref{tb_BT_conformations}. 
The main objectives are:
\begin{enumerate}\setlength{\itemsep}{-0.39ex}
\item
Under Plan A: replicate experiments to achieve the same or better target energy values published 
for standard instances with chains of length of 20, 24, and 
25, given a fixed binary 
coordinate~\cite{Lib-OPUS-protein-folding-1993-JMB-Unger},~\cite{Lib-OPUS-protein-folding-2005-GECCO-Bui}. 
Return  solutions as ternary coordinates.
\item
Under the first Plan C: find {\em simultaneously}, the pair of binary {\em and} ternary coordinates
that maintain the weight of binary coordinates under 
%the 
Plan A -- at the same or better target energy values.
\item
Under the second Plan C: find {\em simultaneously}, the pair of binary {\em and} ternary coordinates 
that maintain the weight of binary coordinates under 
%the 
Plan A and exceed the
energy target value found under the first experiments of 
%the 
Plan C.
\end{enumerate}

\hspace*{-1.3em}
Our findings so far:
\begin{itemize}\setlength{\itemsep}{-0.39ex}
\item
Under Plan A, our experiments replicate but not improve published energy target values.
\item
Under the first Plan C, our experiments generate up to 990 (out of 1000 initiated) new and unique solutions. In most cases, energy values remain the same as for Plan A.
However, there also are improvements that lead to experiments under the second Plan C.
\item
Under the second Plan C, experiments with improved energy targets
generate from 11 to 875 unique solutions with the assigned energy target value, except for the chain of length 24, where {\em again}, an improved energy target value is observed for two instances.
\end{itemize}
\begin{table*}[]
\begin{center}
\caption{
Statistical summary of experiments, a companion to Figure \ref{fg_BT_conformations}.
Experiments under `Referenced solutions' are under Plan A as defined in the paper. 
Experiments under `Equivalent SAW solutions' and `Better SAW solutions' are under Plan C.
All experiments,  except for the one flagged in the footnote, are based on a sample size of 1000.
As an additional bonus, we also found another improved solution while running the case of
L = 24, weight = 10, energy = 10. The energy improved from -10 to -11 and it is this conformation which shown in Figure \ref{fg_BT_conformations}.
}
\label{tb_BT_conformations}
\small
\vspace*{-0.1ex}
\begin{tabular}{c@{~~}|c@{~~}c@{~~}c@{~~}c@{~~}|c@{~~}c@{~~}c@{~~}c@{~~}|c@{~~}c@{~~}c@{~~}c@{~~}|}
%\begin{tabular}{l@{~~}l@{~~~~}l@{}} 

\multicolumn{1}{c}{\textbf{}} & \multicolumn{4}{c}{\textbf{Reference solutions}} & \multicolumn{4}{c}{\textbf{Equivalent SAW solutions}}  & \multicolumn{4}{c}{\textbf{Better SAW  solutions}} \\[0.25ex]
\cline{2-13}
         & energy & \multicolumn{3}{c|}{\textbf{walkLength}} & energy & \multicolumn{3}{c|}{\textbf{walkLength}} & energy & \multicolumn{3}{c|}{\textbf{walkLength}} \\
instance & target & median & mean & stdev & target & median & mean & stdev & target & median & mean & stdev \\
\hline
length = 20 &       &          &         &          &         &         &         &         &        &         &         &       \\
weight = 10 & -9    & 2079	  & 3097 	& 2890.8  &   -9    & 315	  & 812	  & 1183.0 &    -10  & 9742	  & 15088	  & 16364 \\
unique sols & 4     &         &          &         &    928   &         &         &         &    109    &         &         &       \\
\hline  
length = 24 &       &          &         &          &         &         &         &         &        &         &         &       \\
weight = 10 & -9    & 957	  & 1575 	  & 1765  &   -9    & 649	  & 4104	  & 54529 &    -10  & 9059	  & 19391	  & 26142 \\
unique sols & 35    &         &          &         &    975   &         &         &         &     875   &         &         &       \\
\hline       
length = 25 &       &          &         &          &         &         &         &         &        &         &         &       \\
weight = 9  & -8    & 13099 	  & 21557	  & 27639  &   -8    & 959.5	  & 9679	  & 95154 &    -10  & 933928	  & 1243210	  & 1087628 \\
unique sols & 32    &         &          &         &    990   &         &         &         &  11$\dagger$   &         &         &       \\
\hline               
\end{tabular}

\par\vspace*{2.75ex}
\begin{minipage}[t]{0.80\textwidth}
{\em
($\dagger$)
Statistics for the pair (weight = 9, energy target = -10) are based on the sample size of 62 \\
(rather than 1000) as is the case with 
other entries in this table.
}
\end{minipage}

%\end{tiny}
\end{center}
\vspace*{-4.0ex}
\end{table*}

 \section{Conclusions And Future Work}
\label{sec_conclusions}
\vspace*{-1.5ex}
Our experiments 
with the SAW solver
raise the expectation that the
solution of the protein folding problem, where the
chain configuration {\em and} its confirmation
are optimized {\em simultaneously},
may be feasible at an acceptable cost. %at an acceptable cost of computation.
One of the best way to accelerate improvements 
is cooperate with other researchers so 
that solver implementations can be
compared side-by-side for their strengths and weaknesses,
following the example 
in~\cite{Lib-OPUS2-labs-2013-arxiv-Boskovic}.

Experiments are being planned also
for triangular and hexagonal grids
in 2- and 3-dimensions.

%\noindent\vspace*{1.5ex}
%{\sc Acknowledgments}. 
\section*{Acknowledgments} 
\vspace*{-1.5ex}
Computations performed in these experiments could not have been accomplished
without the access to and the  support from the NCSU High Performance Computing Services
(http://www.ncsu.edu/itd/hpc/).
Consultations with Dr. Gary Howell and Dr. Eric Sills are gratefully acknowledged.

The final draft of this paper has been improved with suggestions from Dr. Larry Nevin and Dr. Min Chi.
For the encouragement and the extension of the submission deadline I thank Dr. Andrej \v Zemva.

%\noindent
%\lipsum[1] %\par\lipsum[2]  %\par\lipsum[3] % Dummy text
%\newpage
%\section{References}
%\par\noindent 

{\small
\bibliographystyle{unsrt}
\bibliography{xBib-OPUS,xBib-OPUS2}
}
\end{multicols}

\end{document}